\pdfoutput=1
\documentclass[twocolumn,preprintnumbers,amsmath,amssymb]{revtex4}

\usepackage{graphicx}
\usepackage{dcolumn}
\usepackage{bm}
\usepackage{color}

\begin{document}


\title{Pinpointing Cosmic Ray Propagation With The AMS-02 Experiment}

\author{Miguel Pato}
\email{pato@iap.fr}
\affiliation{Dipartimento di Fisica, Universit\`a degli Studi di Padova, via Marzolo 8, I-35131, Padova, Italy}
\affiliation{Institut d'Astrophysique de Paris, 98bis bd Arago, 75014, Paris, France}
\affiliation{Universit\'e Paris Diderot-Paris 7, rue Alice Domon et L\'eonie Duquet 10, 75205, Paris, France}
\author{Dan Hooper}
\email{dhooper@fnal.gov}
\affiliation{Center for Particle Astrophysics, Fermi National Accelerator Laboratory, Batavia, IL  60510}
\affiliation{Department of Astronomy and Astrophysics, University of Chicago, Chicago, IL  60637}
\author{Melanie Simet}
\email{msimet@uchicago.edu}
\affiliation{Department of Astronomy and Astrophysics, University of Chicago, Chicago, IL  60637}

\date{\today}

\begin{abstract}

The Alpha Magnetic Spectrometer (AMS-02), which is scheduled to be deployed onboard the International Space Station later this year, will be capable of measuring the composition and spectra of GeV-TeV cosmic rays with unprecedented precision. In this paper, we study how the projected measurements from AMS-02 of stable secondary-to-primary and unstable ratios (such as boron-to-carbon and beryllium-10-to-beryllium-9) can constrain the models used to describe the propagation of cosmic rays throughout the Milky Way. We find that within the context of fairly simple propagation models, all of the model parameters can be determined with high precision from the projected AMS-02 data. Such measurements are less constraining in more complex scenarios, however, which allow for departures from a power-law form for the diffusion coefficient, for example, or for inhomogeneity or stochasticity in the distribution and chemical abundances of cosmic ray sources.

\end{abstract}

\maketitle

\section{Introduction}\label{secintro}

Despite nearly a century of observational and theoretical progress, the origin of the cosmic ray spectrum remains a major puzzle of modern astrophysics. The task of identifying the sources of these particles is complicated by the non-trivial processes involved in cosmic ray propagation. Although cosmic ray composition and spectrum measurements have taught us a great deal about the acceleration and propagation of cosmic rays, we still lack a detailed and self-consistent understanding of how these particles are produced, and how they travel through and interact with the interstellar medium.

Below the spectral feature known as the knee ($E \sim 10^{15}$ eV), the bulk of the cosmic ray spectrum is believed to be of galactic origin. Non-relativistic shocks occurring in supernova remnants seem to be likely sources, and are predicted to accelerate cosmic rays with a power-law injection spectrum $Q\propto E^{-\gamma}$ with $\gamma\sim 2$ \cite{Longair2}. At higher energies, even less is known about the origin of the cosmic ray spectrum. In this work, we focus solely on galactic cosmic rays at energies well below the knee. 

Our understanding of how cosmic rays propagate through the Milky Way is informed largely by measurements of the spectra of various cosmic ray species as observed at Earth. In particular, by comparing the spectrum of particles produced in cosmic ray accelerators (primaries) to those that are produced by inelastic processes during propagation of primary particles (secondaries), we can learn about the mechanisms involved in cosmic ray propagation. While stable secondary-to-primary ratios (such as boron-to-carbon and antiproton-to-proton) provide information that can be used to constrain the effective column density cosmic rays pass through before reaching the Solar System, unstable ratios (such as beryllium-10-to-beryllium-9) are useful in constraining the time interval since spallation. Combinations of such observations make it possible to constrain the basic properties of relatively simple cosmic ray propagation models. 

To date, some of the most precise cosmic ray measurements over the GeV-TeV energy range have been made by the CREAM (boron-to-carbon)~\cite{CREAM}, PAMELA (antiproton-to-proton)~\cite{PAMELAp}, ISOMAX ($^{10}$Be-to-$^{9}$Be)~\cite{ISOMAX}, and HEAO-3 (Subiron-to-iron, boron-to-carbon)~\cite{HEAO} experiments. These measurements have been used to place fairly stringent constraints on the parameters of the underlying cosmic ray propagation model~\cite{SimetHooper,DiBernardo:2009ku}. In this article, we extend this approach to include data anticipated from the Alpha Magnetic Spectrometer (AMS-02) experiment, which is scheduled to be deployed on the International Space Station in 2010. With its greater acceptance and superior particle identification relative to previous experiments, measurements from AMS-02 are expected to dramatically improve our understanding of the processes involved in galactic cosmic ray propagation.

\section{Cosmic Ray Propagation in the Milky Way}

\par Once injected from their sources into the interstellar medium, charged cosmic rays $-$ unlike photons or neutrinos $-$ undergo a number of processes potentially capable of significantly altering their spectra (for a recent review, see Ref.~\cite{SMPreview}). The Galactic Magnetic Field, in particular, is responsible for deflecting charged particles, leading them to diffuse gradually throughout the Galaxy, following paths resembling a random walk. Particles with greater energy, and therefore rigidity, diffuse more efficiently and tend to escape the Galaxy more quickly, whereas less energetic cosmic rays are typically confined by the Galactic Magnetic Field for a greater duration.  

An essentially inevitable consequence of high energy particle scattering in the turbulent magnetic field is stochastic acceleration, also known as diffusive reacceleration~\cite{HS94}. This mechanism gives rise to diffusion in momentum space with a diffusion coefficient determined by the spatial diffusion coefficient and the Alfv\'en velocity, which represents the typical velocity at which magnetic irregularities propagate in the interstellar medium.

Other potentially important effects to consider include galactic winds, which may result in the convection of particles away from the Galactic Plane, as well as various energy loss processes. Such energy losses occur as a result of the cosmic rays traversing the galactic medium, which is permeated with gas, radiation fields, and magnetic fields. In the case of nuclei Coulomb and ionization energy losses exist, but they play only a minor role in their propagation. On the other hand, GeV electrons and positrons lose significant quantities of energy through inverse Compton and synchrotron processes; at lower energies, ionization, Coulomb interactions, and bremsstrahlung processes may also be relevant. Furthermore, the decays of unstable, radioactive species must be taken into account, including the introduction of any relevant decay products. Lastly, cosmic ray spallation on the interstellar medium can lead to the extinction of the incident particle and to the creation of a secondary flux consisting of gamma rays from neutral pion decay, electrons, positrons, protons, antiprotons and nuclei. The rate at which spallation occurs is fixed by the nuclear cross sections involved, and by the distribution of gas (mainly H and He) present in the Milky Way.

\par The transport equation that describes all of the above-mentioned processes for a cosmic ray species $i$ (with atomic number $Z_i$ and mass number $A_i$) is given by~\cite{SMPreview,Longair2}
\begin{align}\label{mastereq}
\frac{\partial n_i}{\partial t} = & Q_{tot,i}(\textbf{x},p,t)+\vec{\nabla}\cdot\left(D_{xx}(\textbf{x},R)\vec{\nabla}n_i - \vec{V}_c(\textbf{x})n_i\right) \nonumber \\
+& \frac{\partial}{\partial p} p^2 D_{pp}(\textbf{x},R) \frac{\partial}{\partial p} p^2 n_i -\frac{n_i}{\tau_{d,i}}-\frac{n_i}{\tau_{sp,i}} \nonumber \\
-& \frac{\partial}{\partial p}\left(\dot{p}_i(\textbf{x},p)n_i-\frac{p}{3}\,\vec{\nabla}\cdot \vec{V}_c(\textbf{x})\,n_i\right),
\end{align}
where $n_i=n_i(\textbf{x},p,t)\equiv d^2N/dVdp$ is the number density of particles of species $i$ per unit momentum, $R=pc/(Z_i|e|)$ is the rigidity, $D_{xx}(\textbf{x},R)$ is the spatial diffusion coefficient, $\vec{V}_c(\textbf{x})$ is the convection velocity, $D_{pp}(\textbf{x},R)$ is the momentum space diffusion coefficient, $\tau_{d,i}$ is the decay time, $\tau_{sp, i}$ is the spallation time, and $-\dot{p}_i(\textbf{x},p)$ is the energy loss rate. Notice that the source term $Q_{tot,i}$ includes the primary injection spectrum $Q_{inj}\propto R^{-\gamma}$ (typically assumed to be from acceleration in supernova remnants) as well as the products of decay and spallation of heavier cosmic ray species. $Q_{tot,i}$ may also include exotic primary contributions produced, for example, by dark matter annihilations taking place in the halo of the Milky Way. Such components are not considered in this study, however.

\par The standard approach used to solve Eq.~\eqref{mastereq} is to assume that the cosmic rays are in steady state, such that $\partial n_i/\partial t=0$ (or to solve iteratively the full equation until an approximate steady state is reached). To further simplify the problem, it is common to adopt a cylindrical diffusive region (of radius $r_{max}$ and half-thickness $L$) inside which the diffusion coefficient does not vary with location, and such that the number density of cosmic rays approaches zero at the boundaries. The half-thickness of the cylindrical diffusive region, $L$, is generally much larger than the half-thickness of the galactic disk $h\sim 0.1$ kpc.

To proceed, one may either apply (semi-)analytical methods (e.g.~\cite{analy1,analy2,analy3}) or make use of numerical codes (e.g.~GALPROP \cite{galpropsite,SM98} or DRAGON \cite{dragon1,DiBernardo:2009ku}). In the present work, we use GALPROP v50.1p~\cite{galpropsite} which solves Eq.~\eqref{mastereq} by assuming a homogeneous power-law coefficient, 
\begin{equation*}
D_{xx}(R)=(v/c) D_{0xx}(R/R_0)^{\alpha} ,
\end{equation*}
in a cylindrical diffusive region as described above where the Sun sits at $\textbf{x}_{\odot}=(r_{\odot},z_{\odot})=(8.5,0)$ kpc. GALPROP implements a realistic interstellar hydrogen distribution and interstellar radiation field based on state-of-the-art surveys, and makes use of an up-to-date nuclear reaction network. The primary sources of cosmic rays are assumed to have a unique single or double power-law injection spectrum and isotope composition. Moreover, the distribution of sources can be specified but is typically chosen such that the EGRET gamma ray data are well reproduced. Convection, if implemented, is assumed to move particles along the vertical direction, perpendicular to and away from the disk and with a linear profile,
\begin{equation*}
\vec{V}_c(\textbf{x})=sgn(z)(V_{c,0}+|z|dV_c/dz)\vec{e}_z .
\end{equation*}
Diffusive reacceleration with a given Alfv\'en velocity, $v_A$, is also included, and is modelled with a diffusion coefficient
\begin{equation*}
D_{pp}(R)=\frac{4 p^2 v_A^2}{3 \alpha(4-\alpha^2)(4-\alpha)D_{xx}(R)} .
\end{equation*}
The local interstellar flux of species $i$, $\Phi_i$, follows from the solution of Eq.~\eqref{mastereq} through
\begin{equation*}
\Phi_i(\textbf{x}_{\odot},T)=\frac{c A_i}{4\pi}n_i(\textbf{x}_{\odot},p) ,
\end{equation*}
where $T$ is the kinetic energy per nucleon. Finally, this flux is modulated by the solar wind before arriving at the top of the atmosphere, as calculated using the force field approximation~\cite{GA68}.
Further details pertaining to the GALPROP package are described in Refs.~\cite{galpropsite,SM98}.

\section{Prospects For AMS-02}\label{secAMS02}

\par The second and final version of the Alpha Magnetic Spectrometer (AMS-02) \cite{ams02site,Zuccon} is a large acceptance cosmic ray detector scheduled to be placed onboard the International Space Station in 2010. Over its mission duration of at least three years, it will measure with unprecedented statistics and precision the spectrum of cosmic rays over an energy range of approximately 100 MeV to 1 TeV~\cite{Malinin,Alpat2003}. The AMS-02 instrument will be able to detect and identify nuclei as heavy as iron ($Z\lesssim 26$) with rigidity up to 4 TV~\cite{Zuccon}, and separate isotopes of light elements (namely, H, He and Be) over a kinetic energy range of 0.5$-$10 GeV/n \cite{CasausICRC28,Alpat2007,Arruda}. High precision measurements of the ratios B/C and sub-Fe/Fe (D/p, $^3$He/$^{4}$He and $^{10}$Be/$^{9}$Be) up to energies of $\sim$1 TeV/n ($\sim$10 GeV/n) are anticipated. Due to a high level of proton rejection, positron and antiproton spectra will also be measured with unprecedented precision~\cite{Brun08,Alpat2007}.

\par The principal goals of the AMS-02 experiment~\cite{Zuccon,Casadei2004} include searches for primordial anti-matter among cosmic ray nuclei and signatures of dark matter annihilations or decays. Additionally, however, AMS-02 will contribute considerably to our understanding of the origin, acceleration, and propagation of galactic cosmic rays. In fact, being the largest acceptance ($\sim 0.45 \textrm{ m}^2\textrm{sr}$) space-based magnetic spectrometer, AMS-02 will bypass the atmospheric systematics which affect baloon-borne experiments and simultaneously feature an acceptance more than two orders of magnitude above the $\sim 0.002 \textrm{ m}^2\textrm{sr}$ of the PAMELA satellite \cite{PAMELAsite}. Besides largely improved statistics, AMS-02 will also provide cosmic ray flux measurements up to $\sim$ TeV/n and separate Be isotopes up to $\sim 10$ GeV/n while PAMELA can only reach a few hundred GeV/n and separate H and He isotopes. Therefore, a precise B/C ratio over a wide energy range and high energy $^{10}$Be/$^{9}$Be measurements will be obtained. These data are extremely useful in constraining cosmic ray diffusion parameters. Of particular importante will be the ratio $^{10}$Be/$^{9}$Be since as of today there are no data points above 3 GeV/n.

\par Focusing on the ratios B/C, $^{10}$Be/$^{9}$Be and $\bar{p}/p$, we can estimate the systematic and statistical errors of AMS-02. To begin, the rigidity resolution of AMS-02 will be $\Delta R / R \sim$$1-2$\% at $\sim$$10$ GV, for both protons and He nuclei, and around 20\% at $\sim$$0.5$ TV ($\sim$$1.0$ TV) for protons (He)~\cite{Lin,Alpat2007,OlivaPhD}. Assuming similar capabilities for heavier nuclei as well, we take a conservative value $\Delta R / R=20$\%, which in the case of relativistic particles translates directly into a kinetic energy resolution $\Delta T / T\simeq 20$\%. Such resolution allows logarithmic bins of width $\log_{10} \frac{T+\Delta T/2}{T-\Delta T/2} \simeq 0.087$, or 11$-$12 bins per decade. In the following we assume 10 bins per decade of kinetic energy regardless of the cosmic ray species.

\par In order to compute the statistical errors associated with the ratio $N_i/N_j$, we need the number of $i$ and $j$ particles detected, $N_i=\epsilon_i \, acc_i \, \Phi_i \, \Delta T \Delta t_i$ (and likewise for $j$), $\epsilon_i$ being the efficiency, $acc_i$ the geometrical acceptance of the instrument, and $\Delta t_i$ the operating time. Then, $\frac{\Delta(N_i/N_j)_{stat}}{N_i/N_j}=1/\sqrt{N_i}+1/\sqrt{N_j}$. The geometrical acceptance is a function of the particle type; we adopt $acc_B=acc_C=acc_{Be}=0.45 \textrm{ m}^2\textrm{sr}$ \cite{Rosier,Zuccon}, $acc_p=0.3 \textrm{ m}^2\textrm{sr}$ \cite{Arruda}, and $acc_{\bar{p}}=0.160 (0.033) \textrm{ m}^2\textrm{sr}$ for $\bar{p}$ momenta $1-16$ ($16-300$) GeV \cite{SpadaPlanck,Malinin,Rosier}. Following Ref.~\cite{OlivaPhD}, we fix $\epsilon_B=\epsilon_C=95$\% and all other efficiencies to 100\%. Lastly, we consider one year of operation.

\par As for systematics, we also estimate the errors associated with the mismeasurement of the atomic number of cosmic ray nuclei. Using the full capabilities of the AMS-02 silicon tracker, the author of Ref.~\cite{OlivaPhD} used Monte Carlo simulations to estimate the level of misidentifications, finding fewer than one percent for $2\leq Z \leq 11$. Conservatively, we take $\frac{(\Delta N_B)_{syst}}{N_B}=\frac{(\Delta N_C)_{syst}}{N_C}=1$\%. The $^{10}$Be/$^{9}$Be measurement, on the other hand, is more delicate since it relies not only on charge, but also mass separation. From Ref.~\cite{Pereira}, we take a mean mass resolution for Be of $\Delta m/m\sim 2.5$\%. Requiring a separation of consecutive isotopes within 0.5 atomic mass units, this mass resolution results in misidentification of $^{9}$Be as $^{10}$Be and vice-versa less often than $f\sim 2.275$\%. Hence we use a systematic error for $N_{^{10}Be}$ given by $(\Delta N_{^{10}Be})_{syst}/N_{^{10}Be}=|-f+f\, N_{^{9}Be}/N_{^{10}Be}|$, and analogously for $N_{^{9}Be}$. Clearly, the systematics become unacceptable when either $f\, N_{^{9}Be}/N_{^{10}Be}$ or $f\, N_{^{10}Be}/N_{^{9}Be}$ approaches unity. Finally, for the antiproton-to-proton ratio, the dominant fraction of the systematic error comes from the $\bar{p}$ measurement. In order to confidently identify antiprotons, the large background of protons and electrons must be rejected with high efficiency. In the multi-GeV energy range, at which $p/\bar{p}\sim 10^{4}$ ($e^{-}/\bar{p}\sim 10^2$)~\cite{PAMELAp,SpadaPlanck}, the rejection power expected for AMS-02 is $p:\bar{p}\sim 10^{5}-10^{6}$ ($e^{-}:\bar{p}\sim 10^{3}-10^{4}$)~\cite{SpadaPlanck,Rosier}, leading to a systematic error of order $\frac{10^4}{10^{5}-10^6}=1-10$\% ($\frac{10^2}{10^{3}-10^4}=1-10$\%). Consequently, we take $\frac{(\Delta N_{\bar{p}})_{syst}}{N_{\bar{p}}}=5\%$. We add all of the systematic errors discussed in this paragraph in quadrature to the statistical uncertainties.

\begin{figure}
 \centering
 \includegraphics[width=6.5cm,height=8.5cm,angle=90]{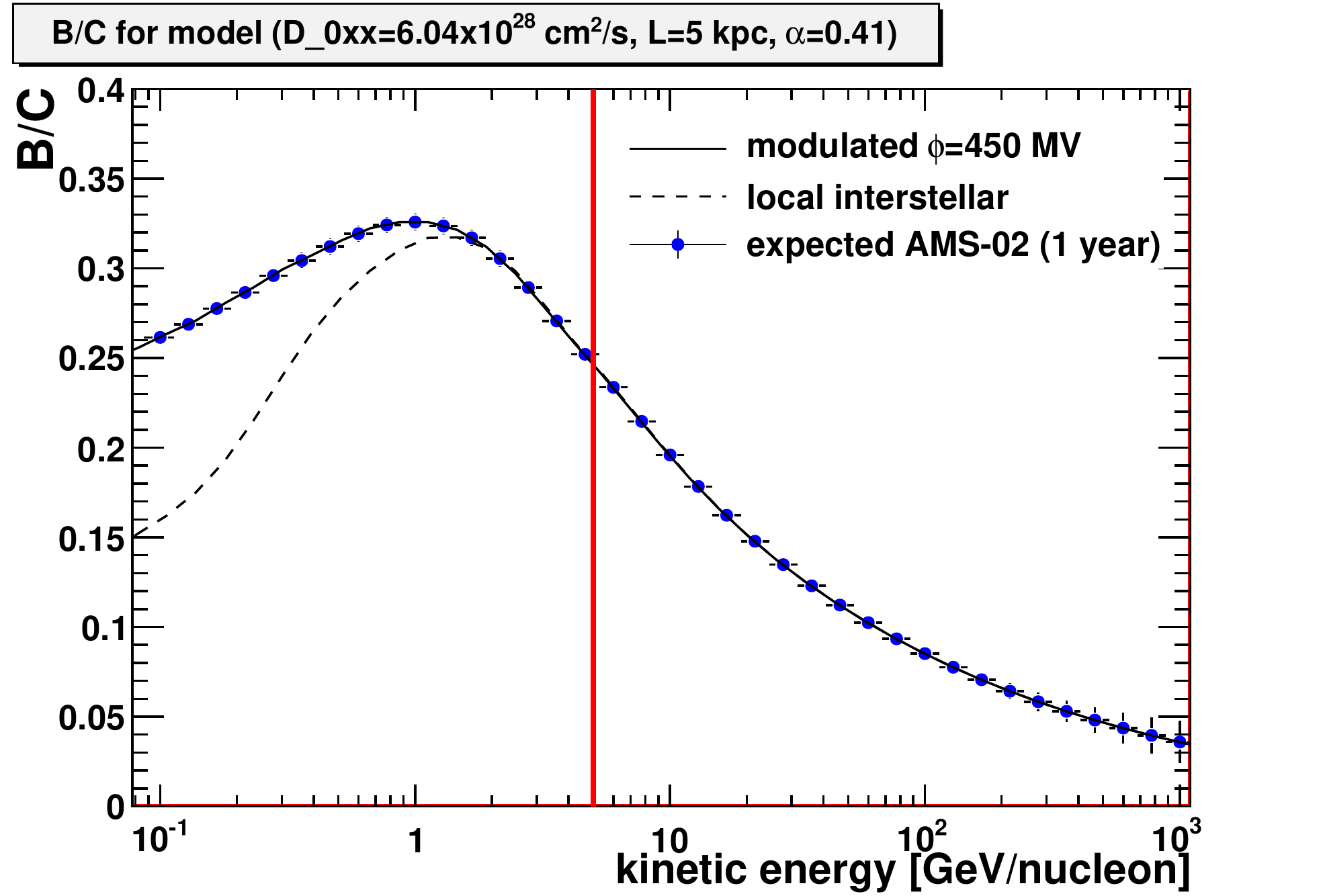}\\
 \includegraphics[width=6.5cm,height=8.5cm,angle=90]{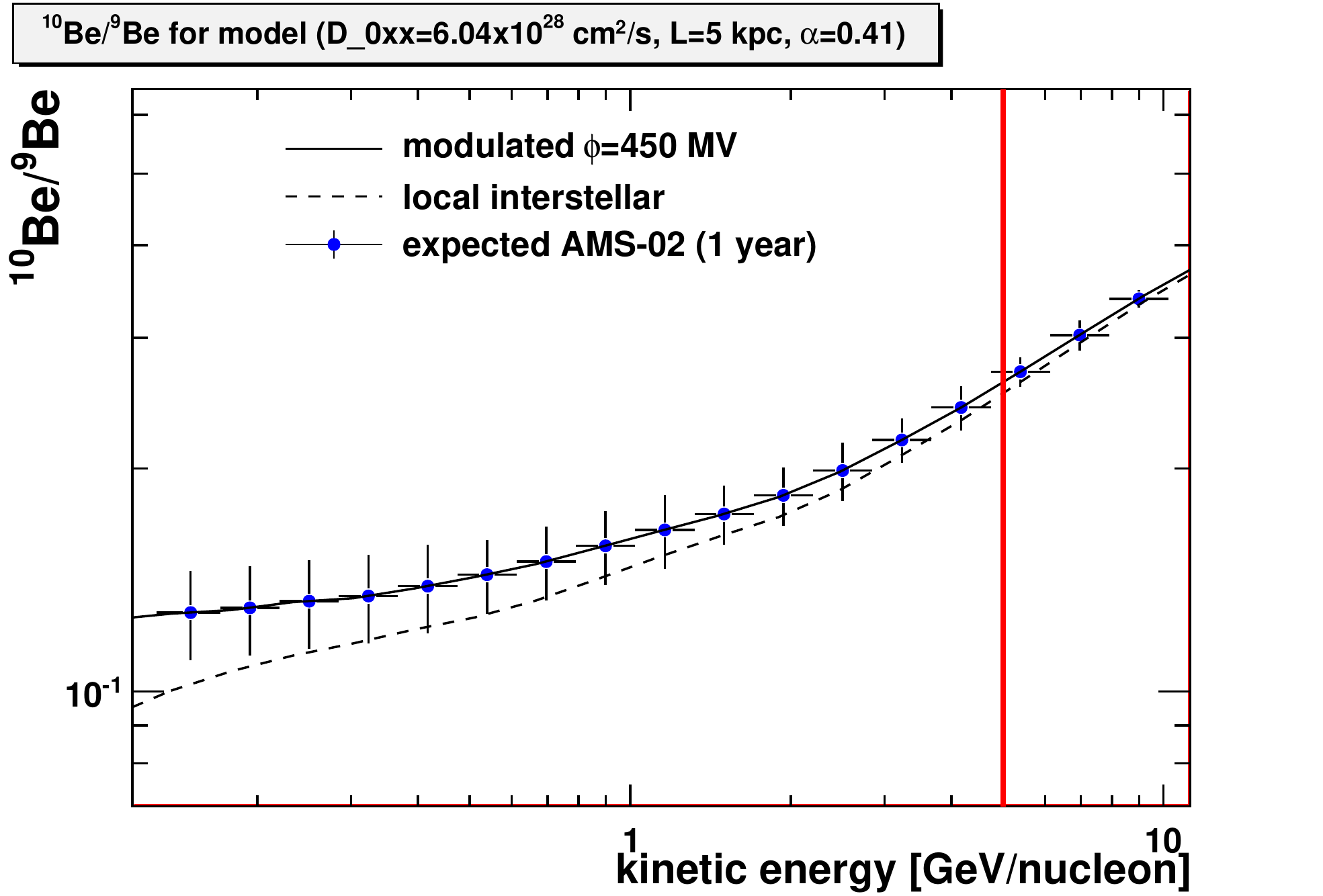}\\
 \includegraphics[width=6.5cm,height=8.5cm,angle=90]{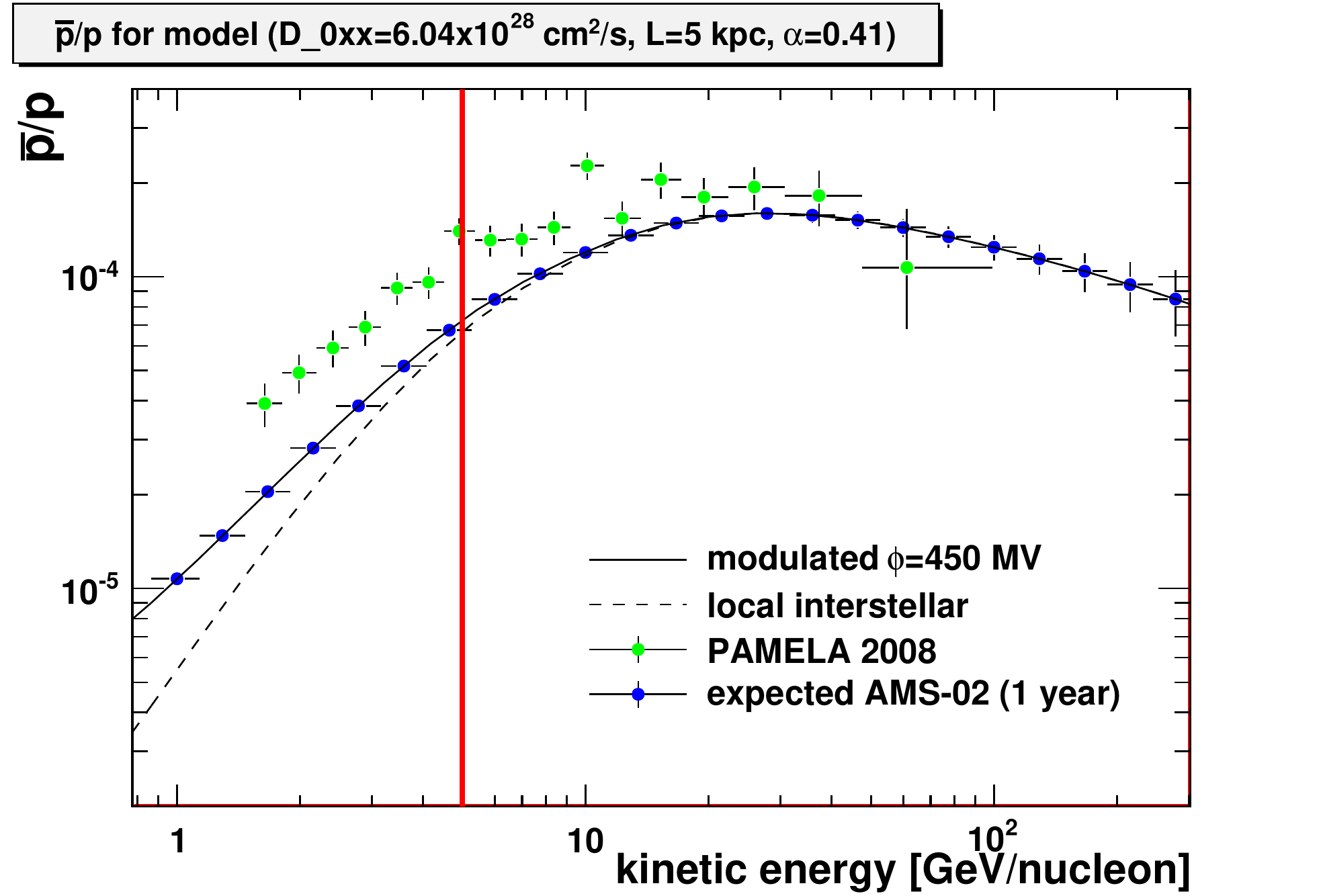}
 \caption{\fontsize{9}{9}\selectfont Projections for ability of the AMS-02 experiment to measure selected stable secondary-to-primary and unstable cosmic ray ratios: B/C, $^{10}$Be/$^{9}$Be and $\bar{p}/p$. As a benchmark, we have considered the best-fit model of Ref.~\cite{SimetHooper}. Systematic and statistical uncertainties are included in the AMS-02 error bars, and 1 year of data taking is assumed. The thick vertical line indicates the energy cut, $T>5$ GeV/n, imposed to reduce the impact of solar modulation on our results. In the lower frame, the PAMELA measurement of the antiproton-to-proton ratio~\cite{PAMELAp} is also shown.}\label{figAMS02}
\end{figure}

\par At this point we can forecast the AMS-02 measurements of various cosmic ray spectra. As a benchmark model, we adopt the best-fit propagation parameters found in Ref.~\cite{SimetHooper}: $D_{0xx}=6.04\cdot 10^{28} \textrm{cm}^2/\textrm{s}$ (at a reference rigidity of $R_0=4$ GV), $L=5$ kpc, $\alpha=0.41$, $v_A=36$ km/s, and no significant convection. The remaining specifications are as in galdef\_50p\_599278 file~\cite{galpropsite}, including a distribution of cosmic ray sources optimized to meet EGRET gamma ray data, and a double power-law injection spectrum $-$ with indices $\gamma_1=1.82$ and $\gamma_2=2.36$ below and above $\tilde{R}_0=9$ GV $-$ to reproduce low-energy cosmic ray data. The ratios B/C, $^{10}$Be/$^{9}$Be and $\bar{p}/p$ corresponding to this model (here after the \emph{true model}) and calculated with GALPROP v50.1p are plotted in Fig.~\ref{figAMS02} along with the projected error bars of the AMS-02 instrument (including both systematic and statistical uncertainties). For detailed simulations of the capabilities of AMS-02, we refer the reader to Ref.~\cite{CasausICRC28,Pereira} and references therein.

\par As shown in the lower frame of Fig.~\ref{figAMS02}, the antiproton-to-proton ratio for the true model is somewhat lower than the values measured by the PAMELA collaboration~\cite{PAMELAp}, especially at low energies. Solar modulation, however, may have a significant impact on this ratio at such energies. As the impact of solar modulation varies with respect to the time period observed, the PAMELA antiproton measurement is not necessarily expected to mimic that to be measured by AMS-02. To reduce the dependence on this effect, we apply an energy cut $T>5$ GeV/n throughout our analysis. In our calculations, we have modulated the cosmic ray spectra with $\phi_F=450$ MV. Common values of the modulation parameter $\phi_F$ range from a few hundred MV up to over a GV (e.g.~\cite{Wied}); however, 450 MV is a reasonable value for data taken around a solar minimum which presumably will be the case of AMS-02 first year. In any case, we stress that our results do not depend much on the solar modulation parameter since only energies above 5 GeV/n are considered. In the remainder of the work, unless otherwise stated, we shall use the modulated data set presented in Fig.~\ref{figAMS02} to perform our analysis.

\section{Constraining propagation models}\label{secconst}

\par In this section, we attempt to estimate how well the projected AMS-02 measurements described in the previous section will be able to constrain the propagation model parameter space. For the moment, we fix the Alfv\'en speed to $v_A=36$ km/s, neglect the effects of convection ($V_{c,0}=dV_c/dz=0$), and proceed in a fashion similar to Ref.~\cite{SimetHooper} to run GALPROP 245 times, in a 7x7x5 grid of the parameters ($D_{0xx},L,\alpha$) over the following ranges: $D_{0xx}=4.54-8.03\cdot 10^{28}\textrm{ cm}^2/\textrm{s}$, $L=3.5-6.5$ kpc and $\alpha=0.39-0.43$. Linearly-(Logarithmically-)spaced gridpoints were implemented for $L$, $\alpha$ ($D_{0xx}$). An infill of 3 points between consecutive gridpoints (corresponding to a reduction of the spacing by a factor 4) was performed and the relevant cosmic ray ratios for each additional propagation model were obtained through 3-dimensional interpolation of the GALPROP runs. The extended grid includes 25x25x17=10,625 different parameter sets. For each set, we calculate the $\chi^2$ using the projected B/C and $^{10}$Be/$^{9}$Be AMS-02 measurements.

\begin{figure}
 \centering
 \includegraphics[width=6.5cm,height=8.5cm,angle=90]{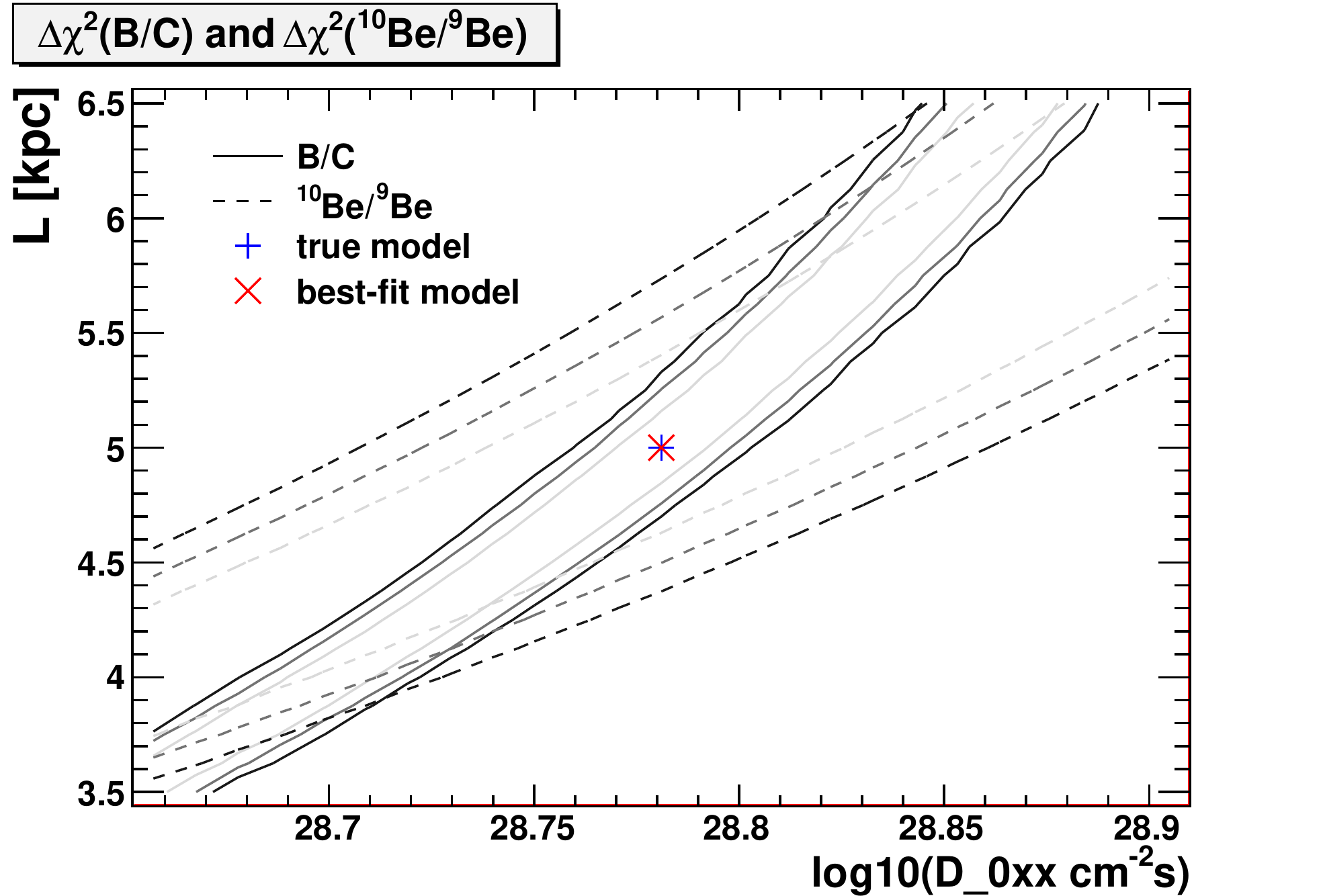}
 \caption{\fontsize{9}{9}\selectfont Constraints on cosmic ray propagation parameters from the projected AMS-02 B/C and $^{10}$Be/$^{9}$Be measurements presented in Fig.~\ref{figAMS02}. The solid lines delimit 1, 2, and 3 $\sigma$ regions using the B/C data set, whereas dashed lines refer to $^{10}$Be/$^{9}$Be. Here, the propagation parameters were varied in the ranges $D_{0xx}=4.54-8.03\cdot 10^{28}\textrm{ cm}^2/\textrm{s}$, $L=3.5-6.5$ kpc and $\alpha=0.39-0.43$. We have assumed $v_A=36$ km/s, $V_{c,0}=dV_c/dz=0$, and have marginalized over $\alpha$.}\label{figBCBe}
\end{figure}

\par Using the above described parameter scan and the projected B/C and $^{10}$Be/$^{9}$Be presented in Fig.~\ref{figAMS02}, we show in Fig.~\ref{figBCBe} the resulting 1, 2 and 3$\sigma$ regions in the $L$ vs.~$D_{0xx}$ plane where we have marginalized over $\alpha$. From this figure, one can immediately identify the complementarity between stable secondary-to-primary and unstable ratio measurements. Whereas stable secondary-to-primary ratios provide an approximate measure of the quantity $L/D_{0xx}$, unstable ratios help to determine $L^2/D_{0xx}$ (for a fixed value of $\alpha$). The combination of B/C and $^{10}$Be/$^{9}$Be measurements can thus provide a determination of both $L$ and $D_{0xx}$. Although the projected $\bar{p}/p$ data, shown in the lower frame of Fig.~\ref{figAMS02}, introduces some additional information into the analysis, it provides a constraint region with a similar shape but broader than that provided by the B/C data. For this reason, we do not include $\bar{p}/p$ in our chi-squares.

\begin{figure}
 \centering
 \includegraphics[width=6.5cm,height=8.5cm,angle=90]{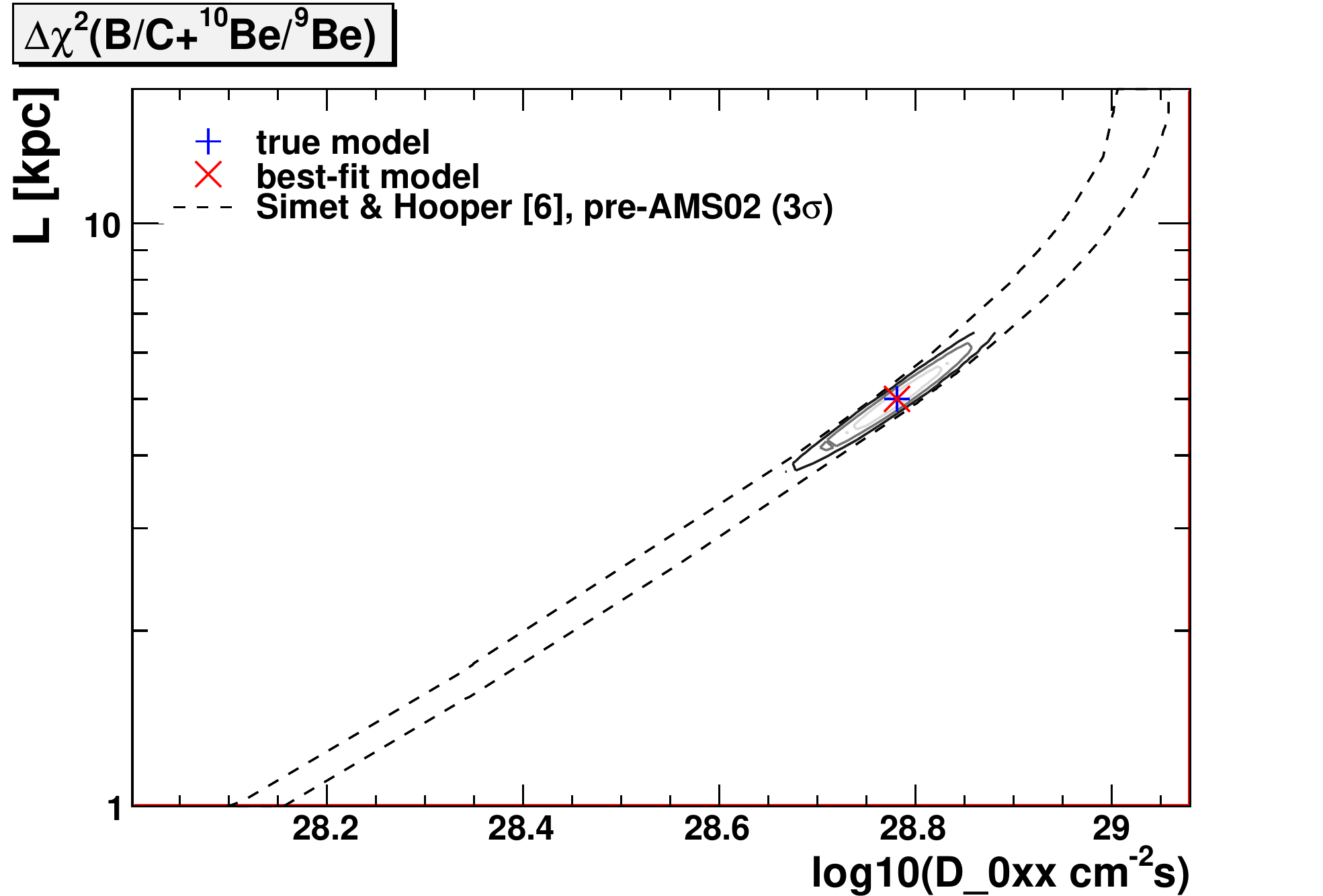}\\
 \includegraphics[width=6.5cm,height=8.5cm,angle=90]{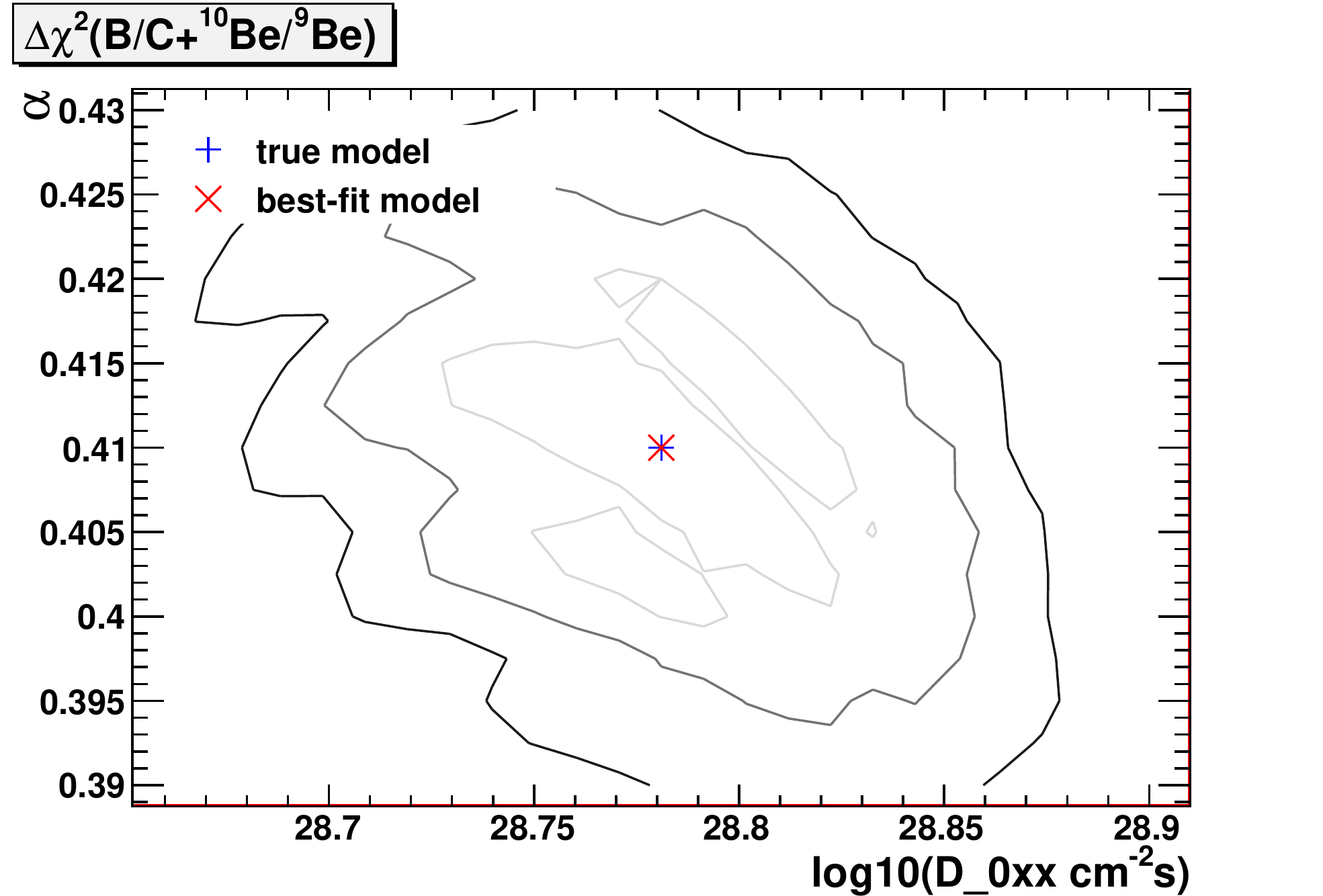}\\
 \includegraphics[width=6.5cm,height=8.5cm,angle=90]{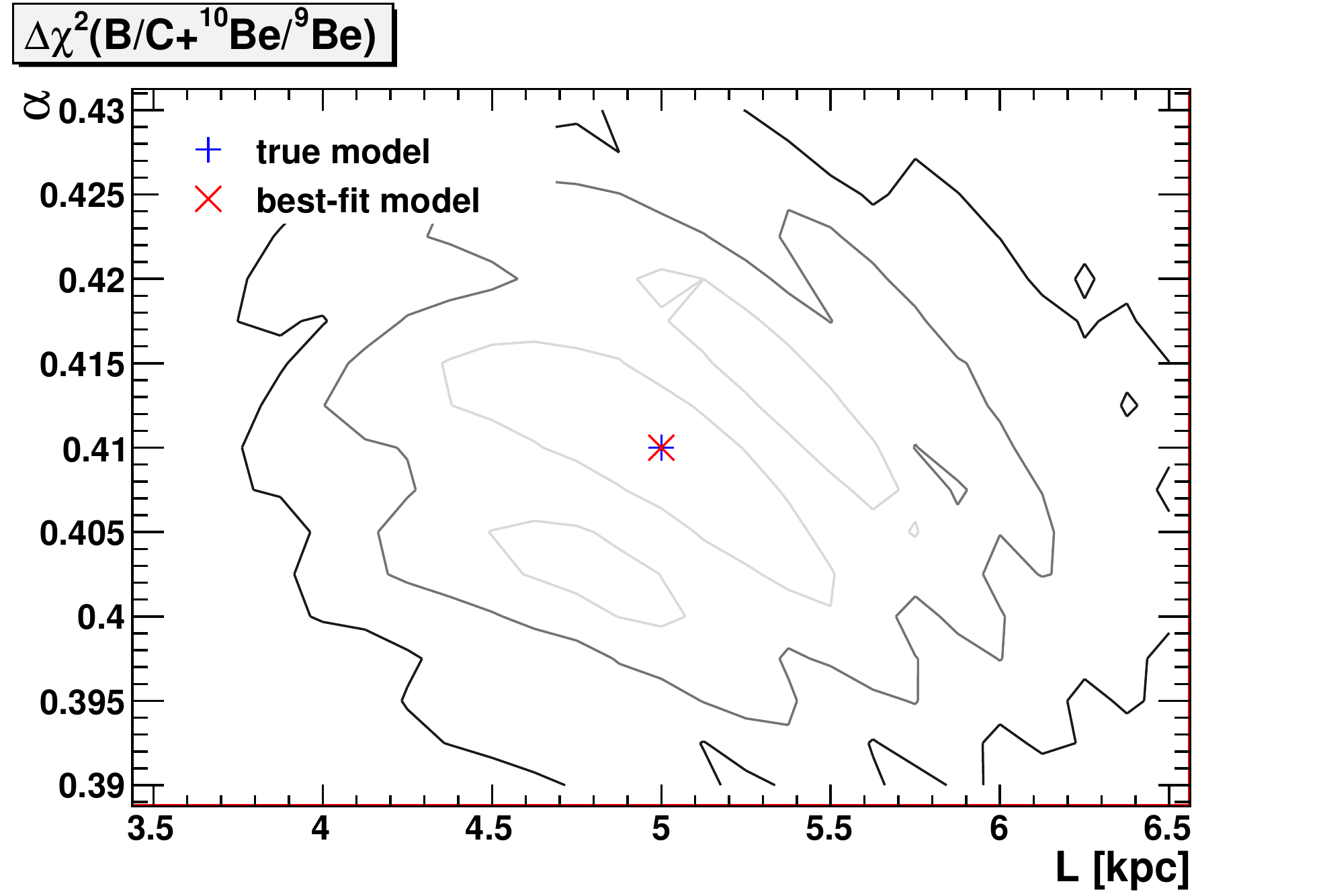}
 \caption{\fontsize{9}{9}\selectfont Regions consistent (within 1, 2 and 3$\sigma$) with projected B/C and $^{10}$Be/$^{9}$Be AMS-02 data from Fig.~\ref{figAMS02} in the $L$ vs.~$D_{0xx}$, $\alpha$ vs.~$D_{0xx}$, and $\alpha$ vs.~$L$ planes. Here, the propagation parameters were varied in the ranges $D_{0xx}=4.54-8.03\cdot 10^{28}\textrm{ cm}^2/\textrm{s}$, $L=3.5-6.5$ kpc and $\alpha=0.39-0.43$. We have assumed $v_A=36$ km/s, $V_{c,0}=dV_c/dz=0$, and have marginalized in each frame over the parameter not shown. In the top frame we show in dashed the 3$\sigma$ contour from Ref.~\cite{SimetHooper}.}\label{figscan1}
\end{figure}

Fig.~\ref{figscan1} shows, for the same scan of propagation parameters as used in Fig.~\ref{figBCBe}, the 1, 2 and 3$\sigma$ contours from the combination of B/C and $^{10}$Be/$^{9}$Be projected measurements presented in Fig.~\ref{figAMS02}. In each frame we have marginalized over the parameter not shown. As this figure demonstrates, the projected AMS-02 measurements of B/C and $^{10}$Be/$^{9}$Be are sufficient (within the context of the simple models presently being considered) to determine the underlying propagation parameters with an accuracy of $\Delta D_{0xx}\sim1.4\cdot 10^{28} \textrm{ cm}^2/\textrm{s}$, $\Delta L \sim1.0$ kpc, and $\Delta \alpha \sim0.02$ (at 1$\sigma$). This precision is much greater than that obtained with present (pre-AMS-02) data; see Refs.~\cite{SimetHooper,DiBernardo:2009ku}. In particular the degeneracy between $D_{0xx}$ and $L$ is broken as can be seen in the upper frame of Fig.~\ref{figscan1} where we have overplotted in dashed the 3$\sigma$ contour from Ref.~\cite{SimetHooper}.

\begin{table}
\centering
\begin{tabular}{c|c|c|c}
\hline
\hline
 \multicolumn{4}{c}{$B/C+^{10}\textrm{Be}/^{9}\textrm{Be}$ \textrm{ best-fit model} \quad $(N_{dof}=21)$ } \\
\hline
 $v_A$ & $dV_c/dz$  &$\quad (D_{0xx} \left[10^{28}\textrm{cm}^2\textrm{s}\right], \, L \left[\textrm{kpc}\right], \, \alpha) \quad$ & $\quad \chi^2/N_{dof} \quad$ \\
\hline
  36 & 0	& $(6.04,\,5.000,\,0.4100)$ & 0 \\
 15 & 0  	& $(6.04,\,8.000,\,0.4850)$ & 1.36 \\
 0 & 0 		& $(6.04,\,8.997,\,0.5000)$ & 2.17 \\
\hline
 0 & 10 	& $(3.89,\,5.623,\,0.5000)$ & 12.78 \\
\hline
\hline
\end{tabular}

\caption{\fontsize{9}{9}\selectfont Best-fit models and reduced chi-squares for the combined B/C+$^{10}$Be/$^{9}$Be projected AMS-02 data set, for various combinations of reacceleration and convection parameters. In each case, the data set used is that presented in Fig.~\ref{figAMS02}. Propagation parameters were varied in the ranges $D_{0xx}=0.57-19.5\cdot 10^{28}\textrm{ cm}^2/\textrm{s}$, $L=1.22-20.48$ kpc and $\alpha=0.32-0.50$. $v_A$ and $dV_c/dz$ are given in units of km/s and km/s/kpc, respectively, and $V_{c,0}=0$. The presence of significant convection will be highly observable to AMS-02. To a lesser extent, the value of the Alfv\'en velocity will also be testable.}\label{tabReaccConv}

\end{table}

\par Thus far, we have not considered any effects of convection and/or variations in the Alfv\'en velocity from our default value of $v_A=36$ km/s. As the quantity and quality of cosmic ray data improves, however, it will become increasingly possible to test these assumptions, and determine the related parameters. In order to study this possibility, we have repeated the procedure described in the previous paragraphs using the following values for the Alfv\'en velocity and the convection velocity: $(v_A [\textrm{km/s}],dV_c/dz [\textrm{km/s/kpc}])=\left\{(36,0),(15,0),(0,0),(0,10)\right\}$, while leaving $V_{c,0}=0$. We have focused here on Alfv\'en velocities smaller than the reference value 36 km/s since some studies (e.g.~Ref.~\cite{DiBernardo:2009ku}) find that lower values of $v_A$ are preferred. For each combination of $v_A$ and $dV_c/dz$, we ran 343 GALPROP jobs in a 7x7x7 grid with ranges $D_{0xx}=0.57-19.5\cdot 10^{28}\textrm{ cm}^2/\textrm{s}$, $L=1.22-20.48$ kpc and $\alpha=0.32-0.50$. Linearly-(Logarithmically-)spaced gridpoints were implemented for $\alpha$ ($D_{0xx}$, $L$). An infill of 3 points between consecutive gridpoints was applied resulting in 25x25x25=15,625 different models. Using the projected B/C and $^{10}$Be/$^{9}$Be data sets shown in Fig.~\ref{figAMS02}, we found the best-fit models given in Table~\ref{tabReaccConv}. As greater departures from our default assumptions are considered, the fits to the projected data become considerably worse.  In particular, even modest ($\sim$$10$ km/s/kpc) amounts of convection lead to very poor fits to the projected data. Large variations in $v_A$ also lead to observable effects, thus enabling AMS-02 to be sensitive to the details of diffusive reacceleration.

\par
\begin{figure}
 \centering
 \includegraphics[width=6.5cm,height=8.5cm,angle=90]{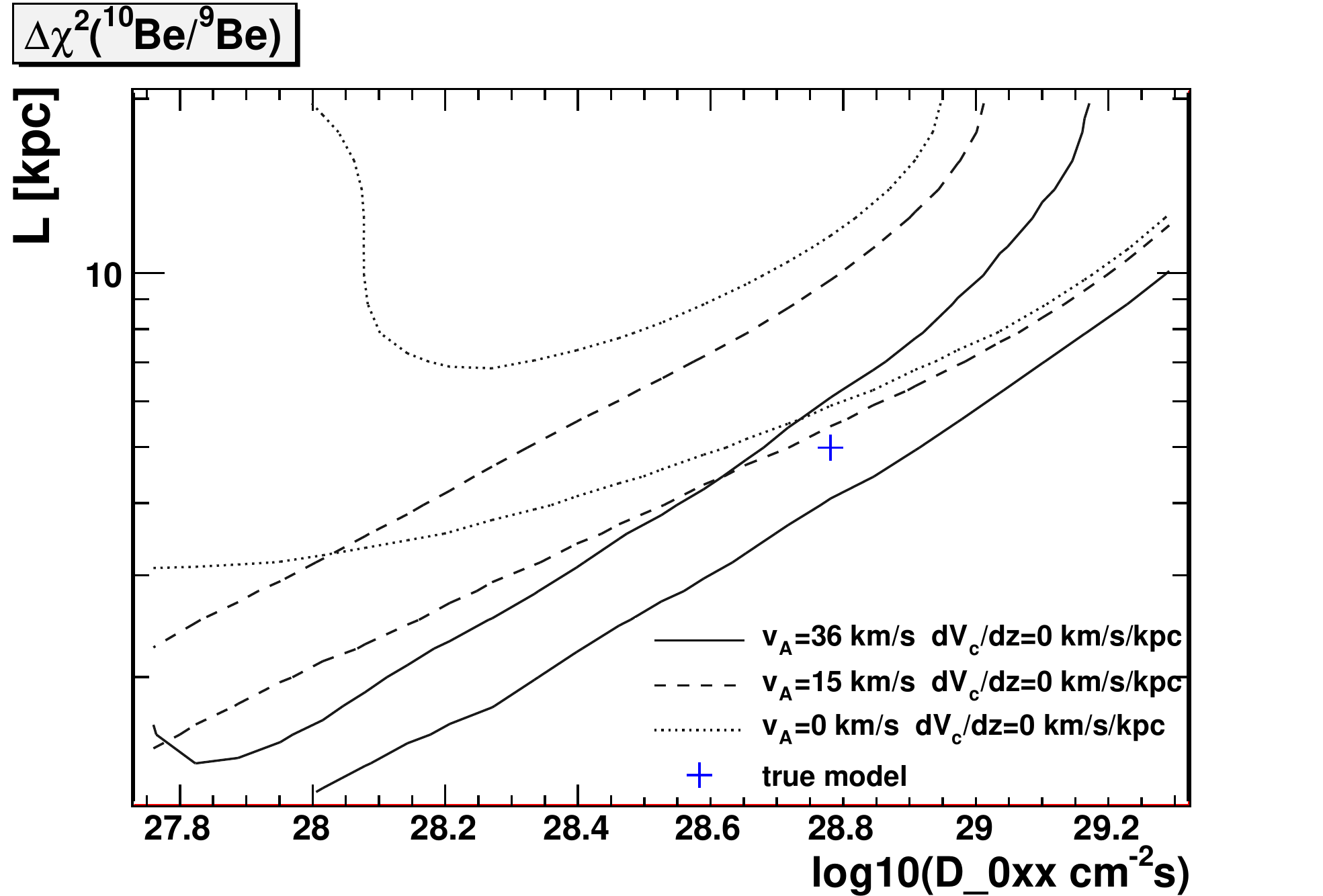}
 \caption{\fontsize{9}{9}\selectfont The effect of changes in the Alfv\'en velocity, $v_A$, in the $^{10}$Be/$^{9}$Be 3$\sigma$ region projected onto the $L$ vs.~$D_{0xx}$ plane. Solid, dashed and dotted lines correspond to $v_A=36,15,0$ km/s, respectively. Propagation parameters were varied in the ranges $D_{0xx}=0.57-19.5\cdot 10^{28}\textrm{ cm}^2/\textrm{s}$, $L=1.22-20.48$ kpc and $\alpha=0.32-0.50$, and the data set used is that presented in Fig.~\ref{figAMS02}. Here, we have neglected convection, and marginalized over $\alpha$. A preference for larger $L$ and lower $D_{0xx}$ is evident as $v_A$ decreases.}\label{figReaccConv}
\end{figure}

 It is well-known that propagation setups with lower Alfv\'en velocities yield lower B/C at energies 1$-$100 GeV/n \cite{SM98,DiBernardo:2009ku}, and that this can be compensated by an increase of $L/D_{0xx}$. But, since reacceleration has negligible influence at high energies, the increase in $L/D_{0xx}$ must be accompanied by a larger value of $\alpha$ so that B/C is sufficiently suppressed in the high energy range. On the other hand, lower values of $v_A$ enhance $^{10}$Be/$^{9}$Be at multi-GeV energies which is also compensated by an increase of $L/D_{0xx}$. This behaviour is illustrated in Fig.~\ref{figReaccConv} where we sketch the 3$\sigma$ contours from the $^{10}$Be/$^{9}$Be projected data shown in Fig.~\ref{figAMS02} for different values of $v_A$. Here, we have used the parameter scan described in the previous paragraph, and marginalized over $\alpha$.

\section{Breaking the assumptions}

\par In order to make the questions addressed in this study tractable, we have thus far relied on a number of simplifying assumptions. In particular, we have assumed homogeneity of the diffusion coefficient $D_{xx}$ over the volume of the diffusion zone, considered cylindrical symmetry for the system, and adopted a smooth distribution of cosmic ray sources with universal injected chemical composition and spectra. While such assumptions are reasonable and have been useful up to this point in time, they will eventually have to be discarded or revised as they break down under the increasing precision of future cosmic ray data. In this section, we study a few of the possible ways in which data from AMS-02 could potentially require us to revise  the assumptions commonly made in modeling galactic cosmic ray production and propagation.

\par
\begin{figure*}
 \centering
 \includegraphics[width=6.5cm,height=8.5cm,angle=90]{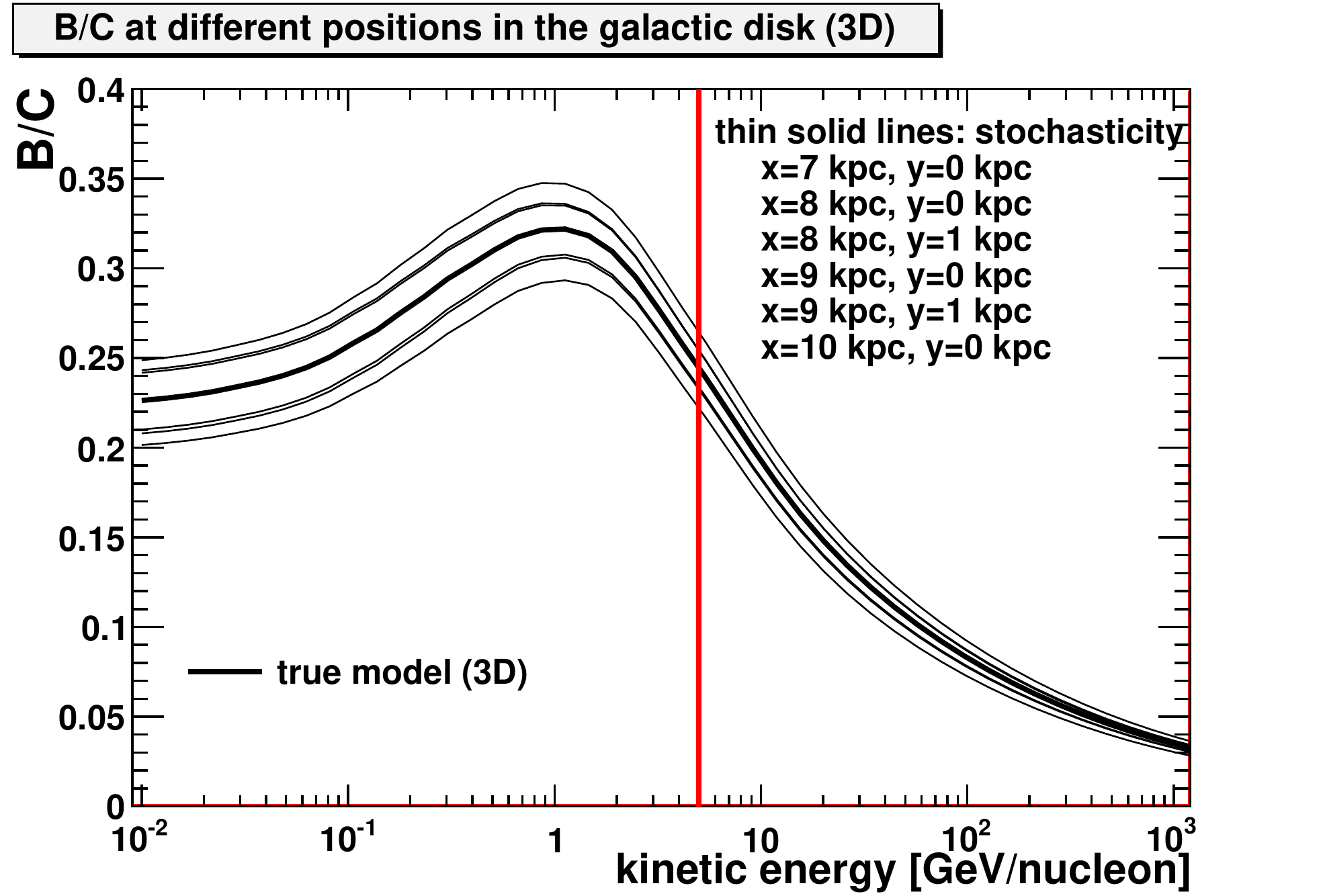}
 \includegraphics[width=6.5cm,height=8.5cm,angle=90]{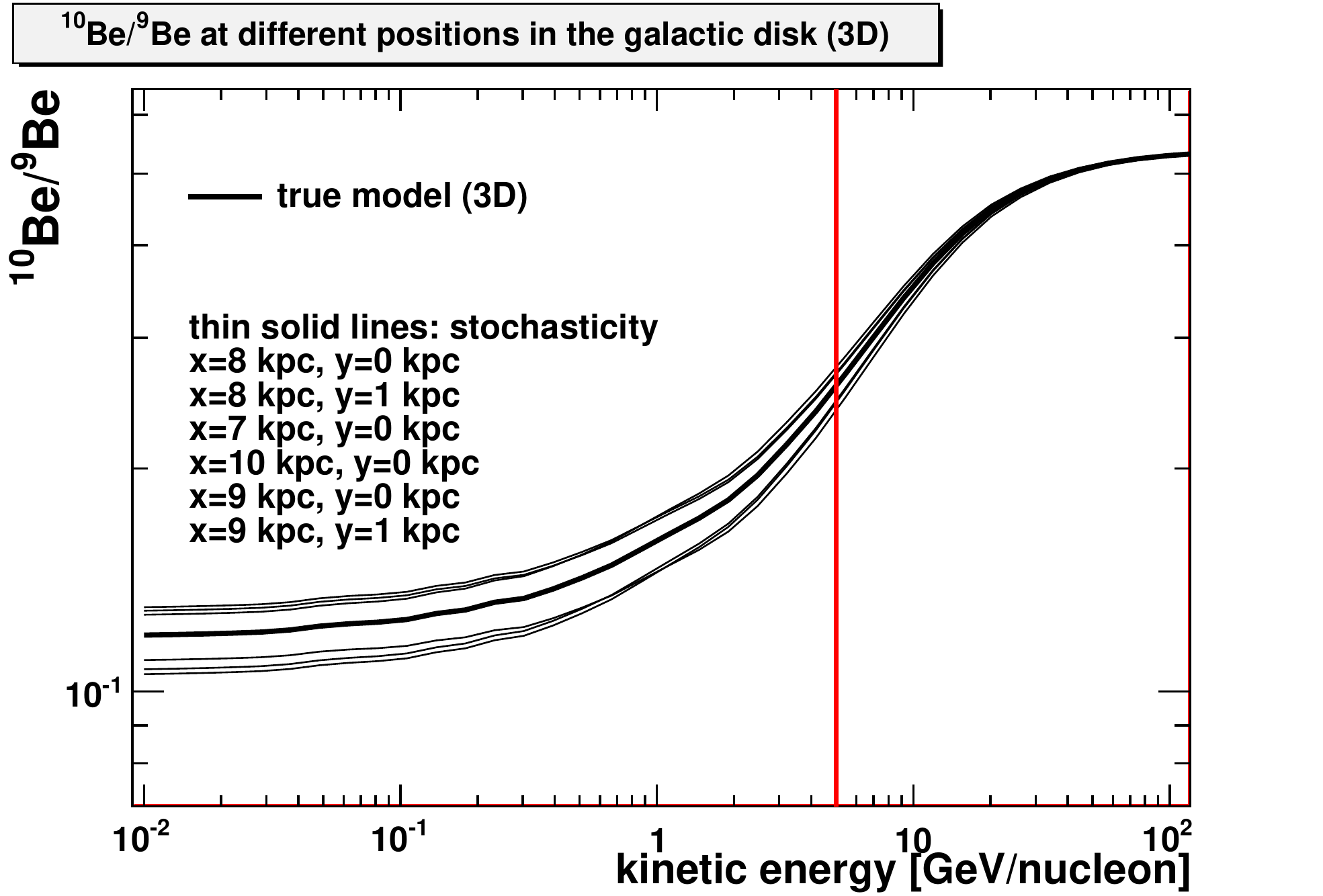}\\
 \includegraphics[width=6.5cm,height=8.5cm,angle=90]{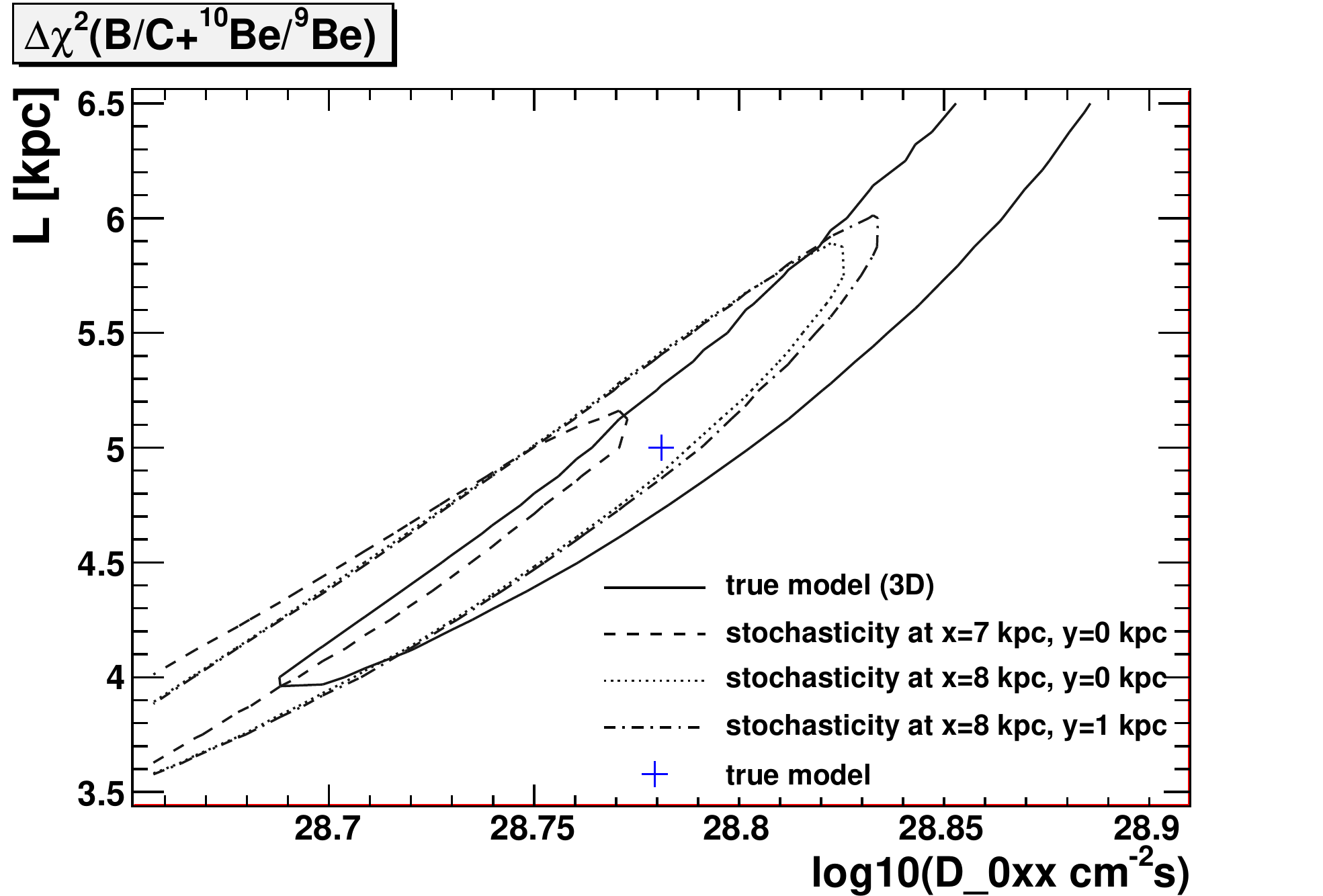}
 \includegraphics[width=6.5cm,height=8.5cm,angle=90]{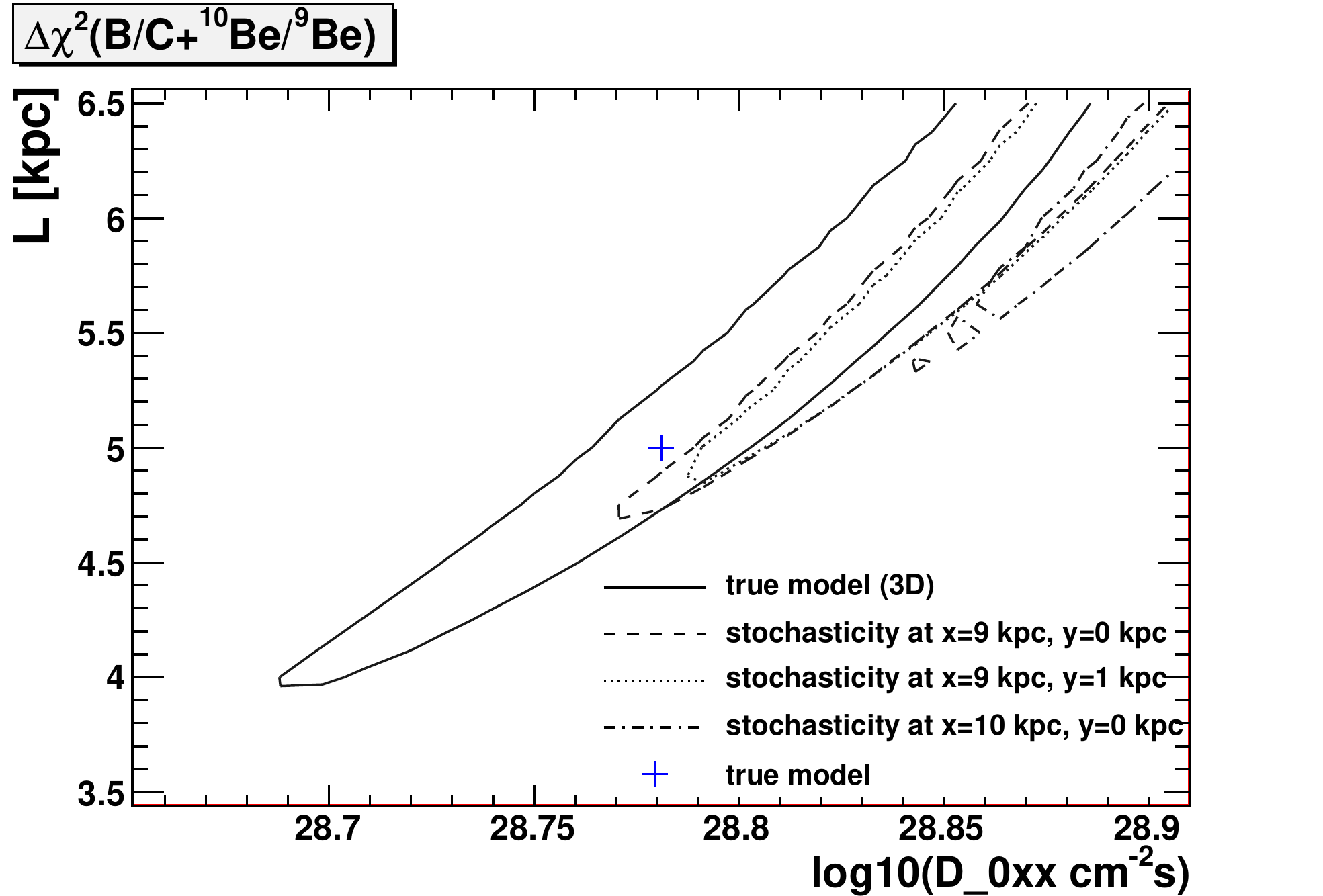}
 \caption{\fontsize{9}{9}\selectfont The impact of the stochastic nature of sources in the injection of cosmic rays on the B/C and $^{10}$Be/$^{9}$Be ratios (top frames), and on the inferred propagation parameters (bottom frames). In the top frames we show the ratios in different positions in the local Galactic Disk $-$ the legend is ordered according to the values of the thin lines at low kinetic energies. In the bottom frames we have marginalized over $\alpha$ and the legend indicates the propagation model used to simulate AMS-02 data; in all cases $v_A=36$ km/s, $V_{c,0}=dV_c/dz=0$, and the propagation parameters were varied in the ranges $D_{0xx}=4.54-8.03\cdot 10^{28}\textrm{ cm}^2/\textrm{s}$, $L=3.5-6.5$ kpc and $\alpha=0.39-0.43$. We have assumed a rate of 0.03 galactic supernova per year and show 3$\sigma$ contours in the lower frames. Stochastic variations limit our ability to deduce the underlying cosmic ray propagation model from stable secondary-to-primary and unstable ratio measurements. See the text for more details.
}\label{sto}
\end{figure*}

 We first consider fluctuations in the recent cosmic ray injection rate. Cosmic ray sources are indeed believed to be of stochastic nature both in space and time $-$ therefore, one does not expect the rate at which cosmic rays are introduced into the Milky Way to be homogeneous or constant over time. Galactic supernova remnants (SNR), in particular, are created at a typical rate of $\sim$ 0.03 per year, and stay active for $\sim 10^4 - 10^5$ yr \cite{Sturner1997}. Depending on whether such an event has taken place recently and nearby, the observed cosmic ray spectrum will vary accordingly. The GALPROP code gives the possibility to model such stochastic fluctuations \cite{galpropsite,ICRC01a,ICRC01b} by solving the transport equation in a three-dimensional spatial grid (unlike thus far used in this work) and defining two further parameters: the average time $t_{SNR}$ between consecutive SNR events occuring in a kpc$^3$ volume around us, and the time interval $t_{CR}$ during which the SNR keeps injecting cosmic rays. We adopt $t_{SNR}=10^4$ yr $-$ corresponding to $\sim$ 0.03 SNR events per year for a standard distribution of sources $-$, $t_{CR}=10^4$ yr and run GALPROP with all other parameters as in the true model of section \ref{secAMS02}. In the upper plots of Fig.~\ref{sto} we show with thin lines the resulting B/C and $^{10}$Be/$^{9}$Be for different positions in the local Galactic Disk. For comparison, the central thick lines denote the ratios obtained at $r=8.5$ kpc in the true propagation model with no stochastic SNR events (but ran with three spatial dimensions). Using the thin lines in the upper frames of Fig.~\ref{sto} to project AMS-02 data, we show in the lower plots how such variations impact the (3$\sigma$) propagation parameter space inferred. Here, $v_A=36$ km/s, $V_{c,0}=dV_c/dz=0$, the propagation parameters were varied in the ranges $D_{0xx}=4.54-8.03\cdot 10^{28}\textrm{ cm}^2/\textrm{s}$, $L=3.5-6.5$ kpc and $\alpha=0.39-0.43$, and we have marginalized over $\alpha$ in the lower frames of Fig.~\ref{sto}. Additionally, we provide in Table~\ref{tabBreak} the best-fit configurations and reduced chi-squares. At some level, stochasticity of cosmic ray sources ultimately limits our ability to deduce the underlying propagation model from stable secondary-to-primary and unstable ratio measurements.

\begin{table*}
\centering
\begin{tabular}{c|c|c|c}
\hline
\hline
 \multicolumn{4}{c}{$B/C+^{10}\textrm{Be}/^{9}\textrm{Be}$ \textrm{ best-fit model} \quad $(N_{dof}=21)$ } \\
\hline
 broken assumption & specification  &$\quad (D_{0xx} \left[10^{28}\textrm{cm}^2\textrm{s}\right], \, L \left[\textrm{kpc}\right], \, \alpha) \quad$ & $\quad \chi^2/N_{dof} \quad$ \\
\hline
\textrm{true model (3D) $^{(*)}$} &                     &  $(6.04,\, 5.000,\, 0.4175)$ & 0.03 [0.06] \\
\hline
  				 &  $x=7$ kpc, $y=0$ kpc & $(4.76,\, 4.000,\, 0.4000)$ & 0.03 [1.49] \\
   				 &  $x=8$ kpc, $y=0$ kpc & $(5.24,\, 4.375,\, 0.4125)$ & 0.02 [0.58] \\
\textrm{Stochasticity $^{(*)}$}	 &  $x=8$ kpc, $y=1$ kpc & $(5.24,\, 4.375,\, 0.4125)$ & $8\cdot10^{-3}$ [0.24] \\
  				 &  $x=9$ kpc, $y=0$ kpc & $(7.48,\, 6.375,\, 0.4275)$ & 0.05 [0.36] \\
  				 &  $x=9$ kpc, $y=1$ kpc & $(7.66,\, 6.500,\, 0.4250)$ & 0.06 [0.55] \\
  				 &  $x=10$ kpc, $y=0$ kpc& $(8.03,\, 6.500,\, 0.4300)$ & 0.41 [2.85] \\
\hline
  		&	$\alpha_1=0.39, \alpha_2=0.43, R_0=4 \textrm{ GV}$       	& $(6.18,\, 5.250,\, 0.4300)$ & 0.07 [0.12] \\
 		&  	$\alpha_1=0.39, \alpha_2=0.43, R_0=10 \textrm{ GV}$   		& $(5.76,\, 5.000,\, 0.4300)$ & 0.03 [0.19] \\
 		&	$\alpha_1=0.39, \alpha_2=0.43, R_0=10^2 \textrm{ GV}$ 		& $(6.04,\, 5.125,\, 0.4000)$ & 0.13 [0.75] \\
 \textrm{Diffusion} &	$\alpha_1=0.39, \alpha_2=0.43, R_0=10^3 \textrm{ GV}$ 		& $(6.04,\, 5.000,\, 0.3900)$ & $8\cdot 10^{-4}$ [$5\cdot 10^{-4}$] \\
 \textrm{Coefficient} &	$\alpha_1=1/3, \alpha_2=1/2, R_0=4 \textrm{ GV}$$^{(+)}$ 	& $(5.21,\, 4.446,\, 0.5000)$ & 0.75 [1.18] \\
 		& 	$\alpha_1=1/3, \alpha_2=1/2, R_0=10 \textrm{ GV}$$^{(+)}$ 	& $(5.21,\, 5.000,\, 0.4775)$ & 0.63 [4.75] \\
 		&	$\alpha_1=1/3, \alpha_2=1/2, R_0=10^2 \textrm{ GV}$$^{(+)}$ 	& $(6.04,\, 5.623,\, 0.3725)$ & 2.45 [16.1] \\
  		&	$\alpha_1=1/3, \alpha_2=1/2, R_0=10^3 \textrm{ GV}$$^{(+)}$ 	& $(6.04,\, 5.000,\, 0.3275)$ & 0.12 [0.07] \\
\hline
\textrm{Source} & 	$^{12}\textrm{C}\times 1.2$       				& $(6.80,\, 5.500,\, 0.4200)$ & 0.02 [0.26] \\
 \textrm{Abundances} &  $^{12}\textrm{C}\times 0.8$       				& $(5.11,\, 4.375,\, 0.3975)$ & 0.04 [0.75] \\
 		&	$(^{12}\textrm{C},^{14}\textrm{N},^{16}\textrm{O})\times 2$     & $(6.18,\, 4.875,\, 0.4125)$ & 0.03 [0.27] \\
\hline
 \textrm{Source} & 	SNR distribution       				& $(5.76,\, 4.625,\, 0.4000)$ & 0.04 [0.05] \\
 \textrm{Distribution} &  pulsar distribution      				& $(5.36,\, 4.750,\, 0.3925)$ & 0.12 [0.39] \\
                  &      reference + nearby source $^{(*)}$                      & $(6.33,\, 5.250,\, 0.4200)$ & 0.03 [0.10] \\        
\hline
\hline
\end{tabular}

\caption{\fontsize{9}{9}\selectfont Best-fit models and reduced chi-squares for the combined B/C+$^{10}$Be/$^{9}$Be data set, assuming different variations to the underlying true model. For each model, an AMS-02 data set was projected according to the corresponding broken assumption. The presented best-fit models were found for $v_A=36$ km/s and $V_{c,0}=dV_c/dz=0$. For configurations marked with $^{(+)}$ the propagation parameters were varied in the ranges $D_{0xx}=0.57-19.5\cdot 10^{28}\textrm{ cm}^2/\textrm{s}$, $L=1.22-20.48$ kpc and $\alpha=0.32-0.50$; for the remaining cases the ranges of the parameter scan were $D_{0xx}=4.54-8.03\cdot 10^{28}\textrm{ cm}^2/\textrm{s}$, $L=3.5-6.5$ kpc and $\alpha=0.39-0.43$. Configurations marked with $^{(*)}$ have been run in GALPROP with three spatial dimensions; for comparison we also present the best-fit model in the case of generating AMS-02 data with the 3-dimensional true model. In the last column the chi-squares in squared brackets were obtained using AMS-02 projected proton flux in addition to the ratios B/C and $^{10}$Be/$^{9}$Be $(N_{dof}=42)$.}\label{tabBreak}

\end{table*}

\par Another example we have considered was the possibility that the diffusion coefficient does not follow a simple power-law, as assumed in the preceding sections. A simple extension of the power-law form is a broken power-law, with indices $\alpha_1$ and $\alpha_2$ below and above a reference rigidity, $R_0$. To explore this possibility, we adopt a model identical to our previous benchmark model (see Sec.~\ref{secAMS02}), but with diffusion coefficient power-law indices of $\alpha_1=0.39$ and $\alpha_2=0.43$ below and above the reference rigidity, $R_0=\left\{4,10,10^2,10^3\right\}$ GV, while fixing $D_{xx}(4\textrm{ GV})=6.04\cdot 10^{28} \textrm{cm}^2/s$. Next, we once again use GALPROP to compute cosmic ray fluxes and ratios in these models and make projections for the observations of AMS-02. Table \ref{tabBreak} shows the results found when we fit these projected data using propagation models without a broken power-law diffusion coefficient. In particular, we use the parameter scan with $v_A=36$ km/s, $V_{c,0}=dV_c/dz=0$, and ranges $D_{0xx}=4.54-8.03\cdot 10^{28}\textrm{ cm}^2/\textrm{s}$, $L=3.5-6.5$ kpc and $\alpha=0.39-0.43$. Disappointingly, good fits are found in each case, thus revealing that small variations in $\alpha$ are unlikely to be discernable from AMS-02 data. In order to verify whether more extreme variations in $\alpha$ are testable, we consider a broken power law for $D_{0xx}$ with indices $\alpha_1=1/3$ and $\alpha_2=1/2$, and $R_0=\left\{4,10,10^2,10^3\right\}$ GV, again fixing $D_{xx}(4\textrm{ GV})=6.04\cdot 10^{28} \textrm{cm}^2/s$. The index 1/3 (1/2) corresponds to a Kolmogorov (Kraichman) power spectrum of magnetic inhomogeneities. Proceeding in the same fashion as before but using the scan of propagation parameters with ranges $D_{0xx}=0.57-19.5\cdot 10^{28}\textrm{ cm}^2/\textrm{s}$, $L=1.22-20.48$ kpc and $\alpha=0.32-0.50$, we obtain the best-fit configurations reported in Table \ref{tabBreak}. Acceptable fits (and wrong best-fit parameters) result in all cases except for $R_0=10^2$ GV. Therefore, if $\alpha$ changes suddenly from 1/3 to 1/2 well inside AMS-02 high energy range $-$ notice that $10^2$ GV corresponds to a proton ($^{10}$B) kinetic energy per nucleon of 99.1 (49.1) GeV/n $-$, the upcoming data on stable secondary-to-primary and unstable ratios should be sufficient to detect the presence of such a drastic break.

\par
\begin{figure}
 \centering
 \includegraphics[width=6.5cm,height=8.5cm,angle=90]{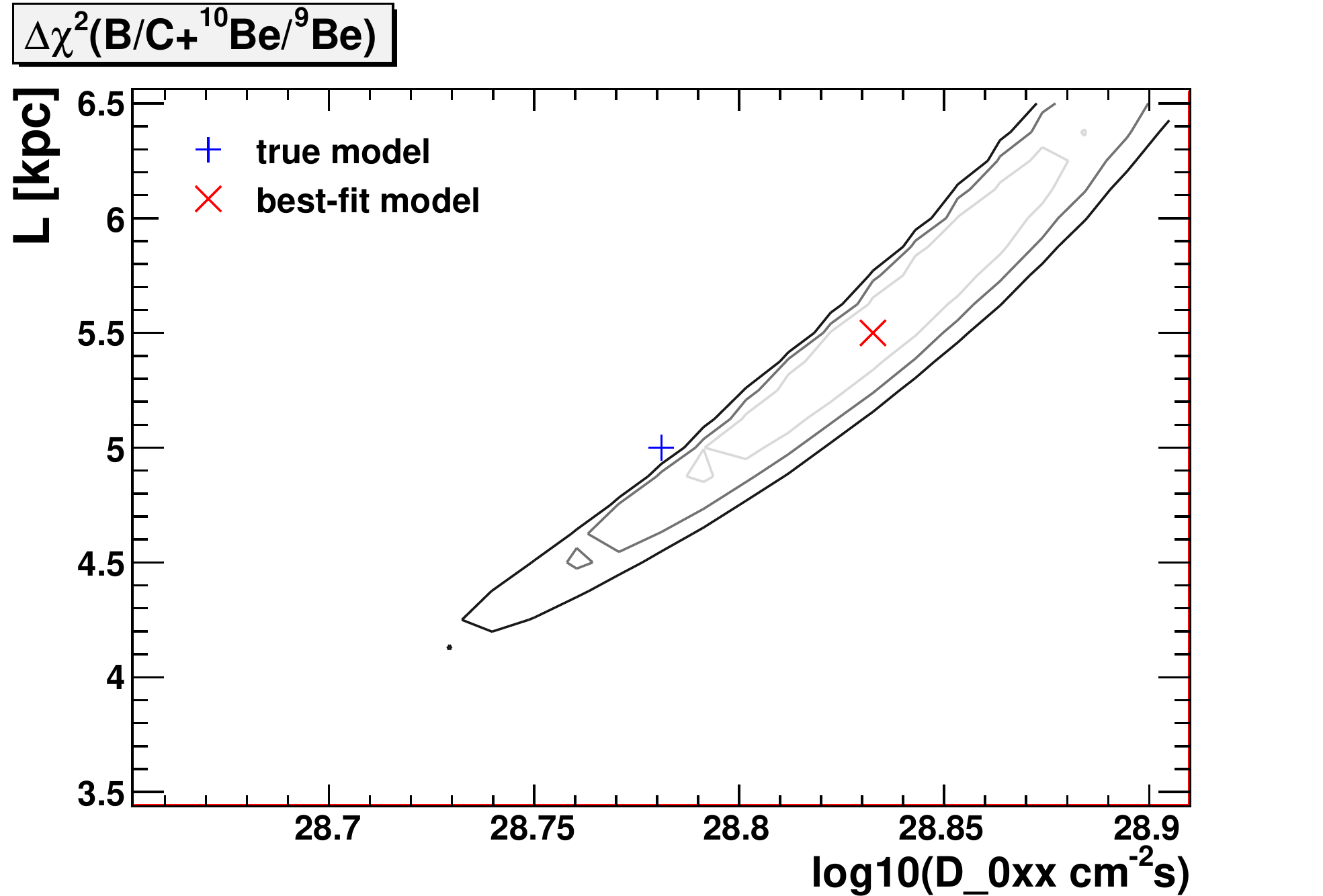}\\
 \includegraphics[width=6.5cm,height=8.5cm,angle=90]{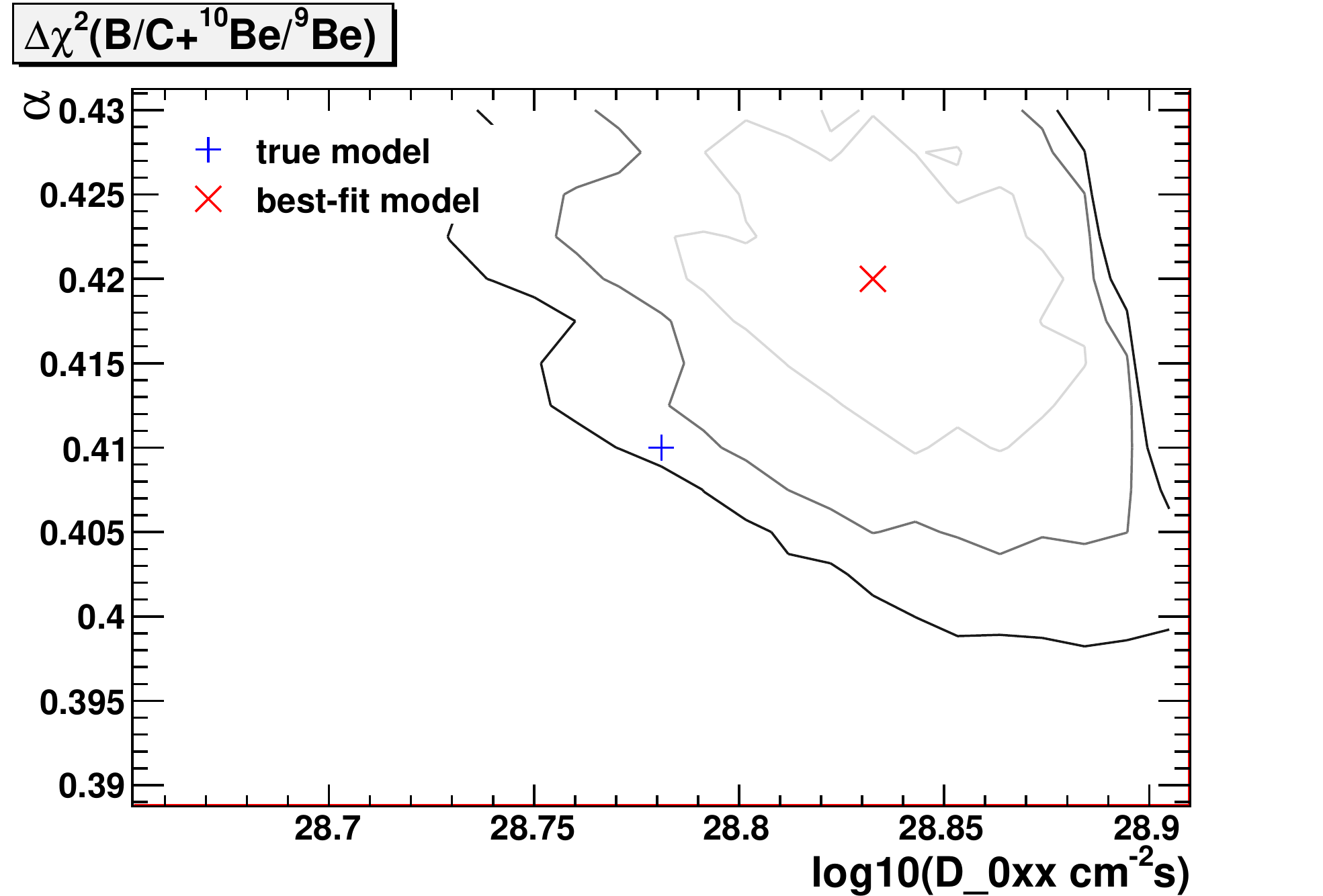}\\
 \includegraphics[width=6.5cm,height=8.5cm,angle=90]{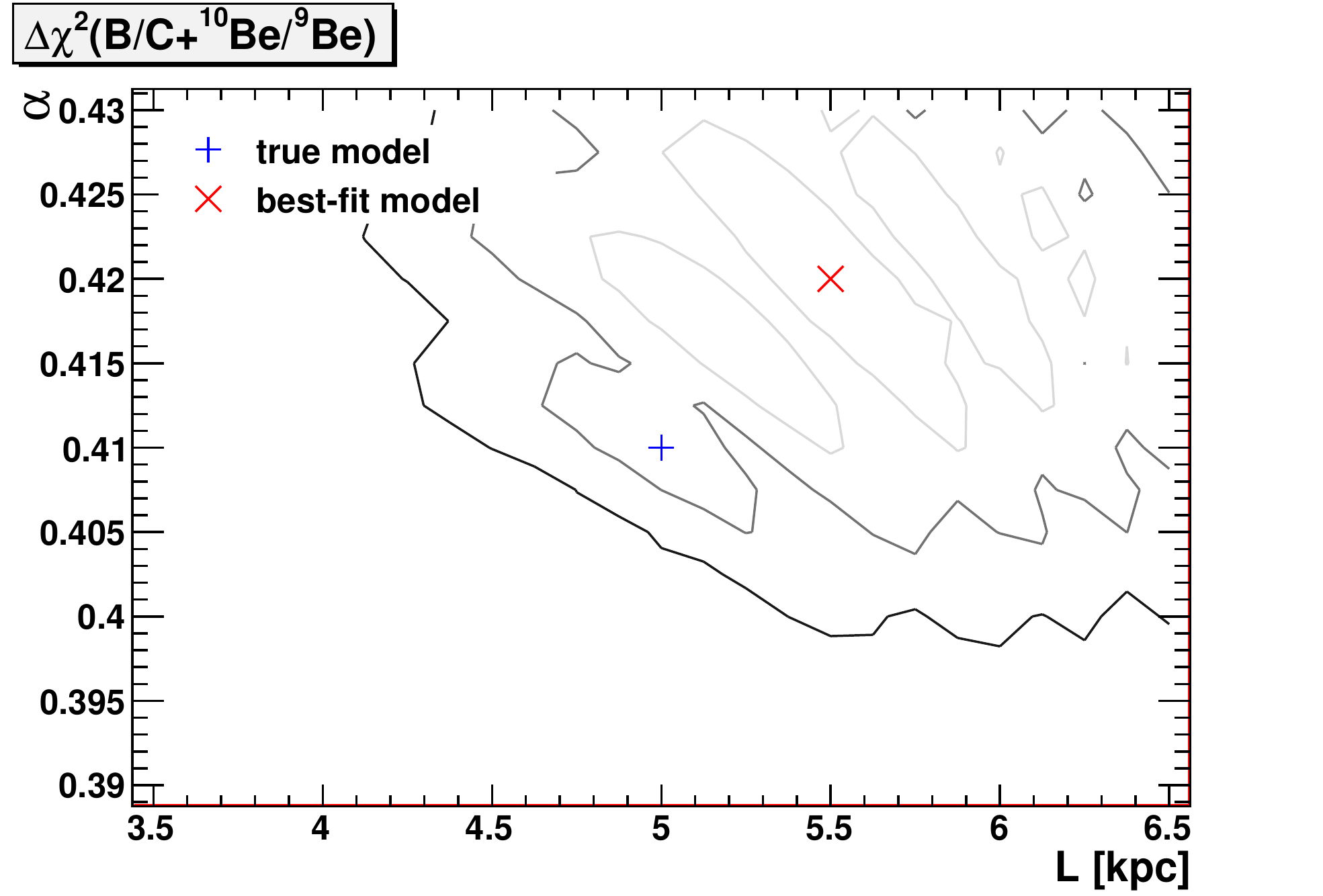}
 \caption{\fontsize{9}{9}\selectfont Regions consistent (within 1, 2 and 3$\sigma$) with projected B/C and $^{10}$Be/$^{9}$Be AMS-02 data in the $L$ vs.~$D_{0xx}$, $\alpha$ vs.~$D_{0xx}$, and $\alpha$ vs.~$L$ planes. Here, the AMS-02 data were projected assuming a true model with a source abundance of $^{12}$C 20\% higher than in the default model of Ref.~\cite{SimetHooper}. Propagation parameters were varied in the ranges $D_{0xx}=4.54-8.03\cdot 10^{28}\textrm{ cm}^2/\textrm{s}$, $L=3.5-6.5$ kpc and $\alpha=0.39-0.43$. We have assumed $v_A=36$ km/s, $V_{c,0}=dV_c/dz=0$, and have marginalized in each frame over the parameter not shown. Notice that, although a good fit was found to the data in this case, the apparent best-fit parameters are significantly different from the values of the true model.}\label{figBreakC}
\end{figure}

 In a similar way, we can test the sensitivity provided by the projected AMS-02 data to the source composition and distribution. A reasonable possibility is that nearby cosmic ray sources produce various species of cosmic rays in relative quantities which differ from the average over the Milky Way (and differ from the default GALPROP assumptions). Unfortunately, Table~\ref{tabBreak} shows that we have very little sensitivity to a $\pm$20\% change in the source abundance of carbon nor even to a factor 2 in the source abundances of C, N and O. In all cases, very good fits to AMS-02 projected B/C and $^{10}$Be/$^{9}$Be are obtained, although the apparent best-fit parameters are not necessarily the parameters of the true model, opening the possibility that we may infer a very well-fit, but incorrect, cosmic ray propagation model from AMS-02 data. We illustrate such effect for the case of a 20\% enhanced source abundance of C: Fig.~\ref{figBreakC} shows the 1, 2 and 3$\sigma$ contours using $v_A=36$ km/s, $V_{c,0}=dV_c/dz=0$, and propagation parameters in the ranges $D_{0xx}=4.54-8.03\cdot 10^{28}\textrm{ cm}^2/\textrm{s}$, $L=3.5-6.5$ kpc and $\alpha=0.39-0.43$. In each frame we have marginalized over the parameter not shown.

\par Moreover, we have studied the impact of varying the distribution of cosmic ray sources. The usual parameterization optimized to reproduce EGRET gamma ray data follows $Q_{inj}\propto (r/r_{\odot})^{\eta}\, \exp(-\xi\frac{r-r_{\odot}}{r_{\odot}}-\frac{|z|}{0.2\textrm{ kpc}})$ with $\eta=0.5$, $\xi=1.0$ and a cut-off radius $r_{max}=20$ kpc \cite{SM98}. Following Ref.~\cite{galpropsite}, we consider alternative scenarios, namely a supernova remnant-like distribution (with $\eta=1.69$, $\xi=3.33$), and a pulsar-like distribution (with $Q_{inj}\propto \cosh(r_{\odot}/r_c)\, \exp(-\frac{|z|}{0.2\textrm{ kpc}}) / \cosh(r/r_c)$, $r_c=3.5$ kpc). As indicated in Table~\ref{tabBreak}, high quality fits are found in both cases; and again the best-fit models do not coincide with the parameters of the true model. The same holds in the case of the usual parameterization with $\eta=0.5$, $\xi=1.0$ and $r_{max}=20$ kpc plus a nearby source that we put at $(x,y,z)=(8.66,0,0)$ kpc (Geminga approximate position). We are, therefore, forced to conclude that AMS-02 data will likely be insensitive to different assumptions pertaining to the source composition and distribution, and that the values of the propagation parameters may even be potentially misinferred as a result.

\par Now, it is interesting to check if any of the best-fit configurations listed in Table \ref{tabBreak} induce cosmic ray fluxes that conflict with the corresponding true model. For such, besides B/C and $^{10}$Be/$^{9}$Be, we also project the one-year AMS-02 proton flux, in the range 5 GeV $-$ 1 TeV and with 3\% of assumed systematics. In the last column of Table \ref{tabBreak} we show in squared brackets the reduced chi-squares of the combined data set including B/C, $^{10}$Be/$^{9}$Be and the proton flux for the best-fit models previously found using B/C and $^{10}$Be/$^{9}$Be only. In some cases $-$ notably when assuming a break from 1/3 to 1/2 in the diffusion coefficient index $-$ the proton flux helps discriminating wrong non-minimal assumptions. In the remaining situations, however, the misinference of propagation parameters discussed in the previous paragraphs still persists.

\begin{table*}
\centering
\begin{tabular}{c|c|c|c|c}
\hline
\hline
 \multicolumn{5}{c}{\textrm{ best-fit model}} \\
\hline
 & \multicolumn{2}{c|}{$B/C+^{10}\textrm{Be}/^{9}\textrm{Be} \, (N_{dof}=21)$} & \multicolumn{2}{c}{$B/C+^{10}\textrm{Be}/^{9}\textrm{Be}+p \, (N_{dof}=42)$} \\
\hline
$\gamma_2$ & $\quad (D_{0xx} \left[10^{28}\textrm{cm}^2\textrm{s}\right], \, L \left[\textrm{kpc}\right], \, \alpha) \quad$ & $\quad \chi^2/N_{dof} \quad$ & $\quad (D_{0xx} \left[10^{28}\textrm{cm}^2\textrm{s}\right], \, L \left[\textrm{kpc}\right], \, \alpha) \quad$ & $\quad \chi^2/N_{dof} \quad$ \\
\hline
 2.26  & $(6.04,\,5.000,\,0.4100)$ & 0.13 &  $(7.00,\,5.000,\,0.3500)$ & 2.55 \\
 2.31  & $(6.04,\,5.000,\,0.4100)$ & 0.03 &  $(7.00,\,5.623,\,0.3875)$ & 0.66 \\
 2.34  & $(6.04,\,5.000,\,0.4100)$ & $5\cdot 10^{-3}$ & $(7.00,\,5.623,\,0.3875)$ & 0.50 \\
 2.38  & $(6.04,\,5.000,\,0.4100)$ & $5\cdot 10^{-3}$ & $(6.04,\,5.000,\,0.4100)$ & 0.56 \\
 2.41  & $(6.04,\,5.000,\,0.4100)$ & 0.03 &  $(6.04,\,5.623,\,0.4475)$ & 1.63 \\
 2.46  & $(6.04,\,5.000,\,0.4100)$ & 0.14 &  $(6.04,\,5.623,\,0.4550)$ & 4.61 \\
\hline
\hline
\end{tabular}

\caption{\fontsize{9}{9}\selectfont Best-fit models and reduced chi-squares for the combined B/C+$^{10}$Be/$^{9}$Be and B/C+$^{10}$Be/$^{9}$Be+p data sets, assuming different high energy injection indices $\gamma_2$. For each model, the AMS-02 data were projected according to the corresponding $\gamma_2$. The presented best-fit models were found for $v_A=36$ km/s, $V_{c,0}=dV_c/dz=0$, $\gamma_1=1.82$, $\gamma_2=2.36$ and $\tilde{R}_0=9$ GV. The propagation parameters were varied in the ranges $D_{0xx}=0.57-19.5\cdot 10^{28}\textrm{ cm}^2/\textrm{s}$, $L=1.22-20.48$ kpc and $\alpha=0.32-0.50$.}\label{tabSpectralIndex}

\end{table*}

\par
\begin{figure}
 \centering
 \includegraphics[width=6.5cm,height=8.5cm,angle=90]{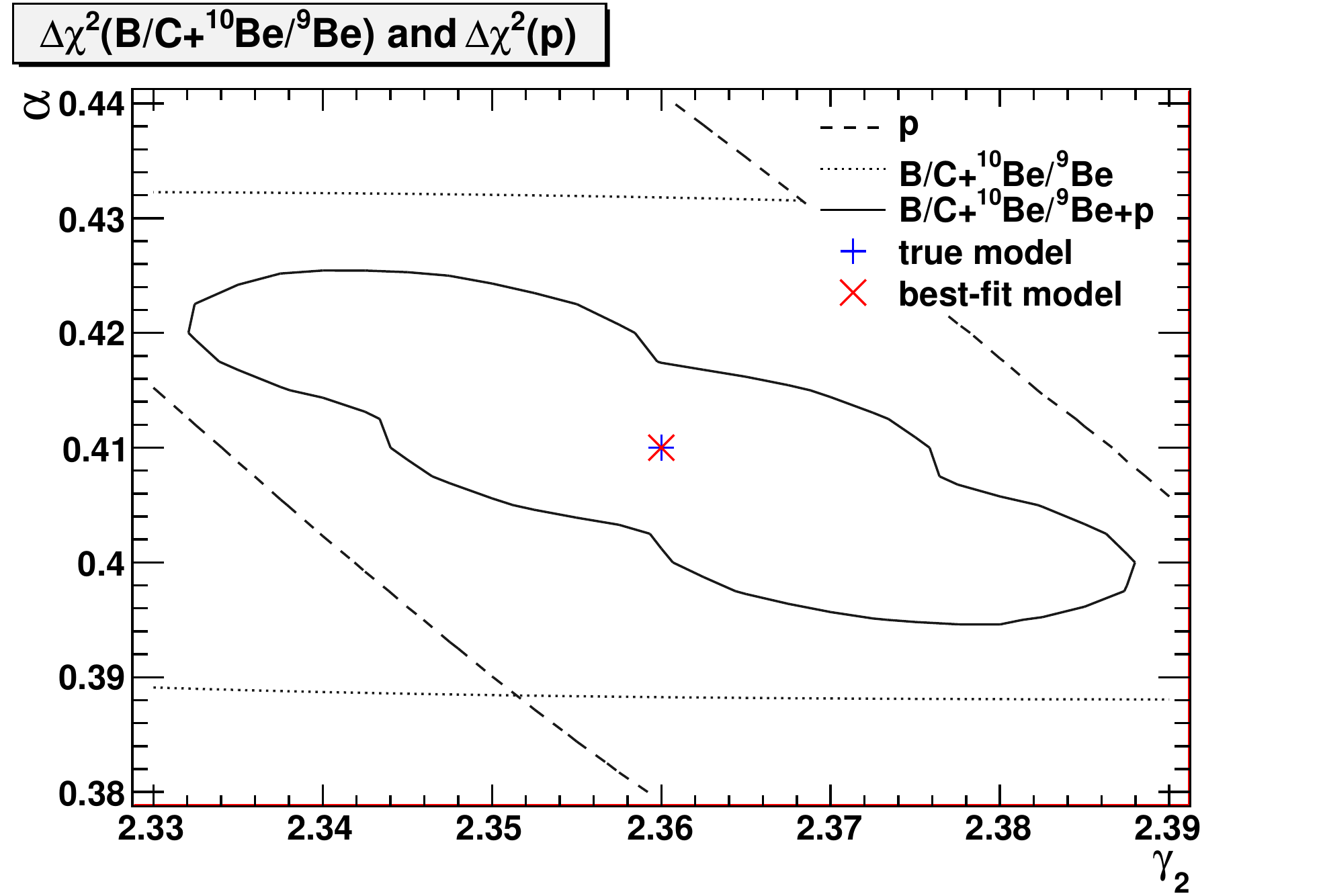}
 \caption{\fontsize{9}{9}\selectfont Inferred parameter regions (within 3$\sigma$) in the plane $\alpha$ vs.~$\gamma_2$. The dashed (dotted) [solid] lines refer to the case where the AMS-02 projected data set consisted of the proton flux (B/C and $^{10}$Be/$^{9}$Be) [proton flux, B/C and $^{10}$Be/$^{9}$Be]. Here, the propagation parameters were varied in the ranges $D_{0xx}=4.54-8.03\cdot 10^{28}\textrm{ cm}^2/\textrm{s}$, $\gamma_2=2.33-2.39$ and $\alpha=0.38-0.44$. We have assumed $v_A=36$ km/s, $V_{c,0}=dV_c/dz=0$, and have marginalized over $D_{0xx}$. The complementarity between the ratios B/C and $^{10}$Be/$^{9}$Be and the proton flux is evident.}\label{figBCBepf}
\end{figure}

 Finally, we turn our attention to the source spectral index. Up to this point the injection spectrum at the sources was assumed to be a double power law in rigidity with indices $\gamma_1=1.82$ and $\gamma_2=2.36$, and a break $\tilde{R}_0=9$ GV as mentioned in section \ref{secAMS02}. We now release this assumption by taking several values for $\gamma_2$ $-$ the low energy rigidity index $\gamma_1$ and the break $\tilde{R}_0$ are kept fixed since we are focussing on high energy data only. Proceeding as earlier, we generate the projected AMS-02 data for each true model and then find the best-fit configuration using only B/C and $^{10}$Be/$^{9}$Be or including also the proton flux $-$ the results are reported in Table \ref{tabSpectralIndex}, where we have used the parameter scan with ranges $D_{0xx}=0.57-19.5\cdot 10^{28}\textrm{ cm}^2/\textrm{s}$, $L=1.22-20.48$ kpc and $\alpha=0.32-0.50$. As can be inferred from this Table, the B/C and $^{10}$Be/$^{9}$Be data set is not sensitive to a variation of $\gamma_2$ in the range 2.26$-$2.46. Such conclusion was expected because cosmic ray ratios are largely independent of the source term. In order to break such degenaracy a cosmic ray flux measurement $-$ e.g.~of protons $-$ must be used. The proton flux at high energies is supposed to go as $E^{-(\alpha+\gamma_2)}$ and consequently it helps discriminating wrong values for the injection index $\gamma_2$. In fact, as shown in Table \ref{tabSpectralIndex}, the analysis with a combined data set including B/C, $^{10}$Be/$^{9}$Be and the proton flux produces relatively large reduced chi-squares of the best-fit configurations. This indicates that $\gamma_2$ is deviating from the assumed value of 2.36. Notice as well that, when one uses the projected proton data, the best-fit values for $\alpha$ are smaller for smaller injection indices $\gamma_2$, and vice-versa $-$ this is because the proton flux is sensitive to the combination $\alpha+\gamma_2$. To proceed further and quantify how sensitive AMS-02 will be to the cosmic ray injection spectrum, we scan the parameter space ($D_{0xx},\gamma_2,\alpha$) and fix $v_A=36$ km/s, $V_{c,0}=dV_c/dz=0$, and $L=5$ kpc (notice that $D_{0xx}$ and $L$ are approximately degenerate). We ran GALPROP 343 times, in a 7x7x7 grid of the parameters ($D_{0xx},\gamma_2,\alpha$) over the ranges $D_{0xx}=4.54-8.03\cdot 10^{28}\textrm{ cm}^2/\textrm{s}$, $\gamma_2=2.33-2.39$ and $\alpha=0.38-0.44$. Linearly-(Logarithmically-)spaced gridpoints were implemented for $\gamma_2$, $\alpha$ ($D_{0xx}$). An infill of 3 points between consecutive gridpoints was applied resulting in 25x25x25=15,625 different models. Figure \ref{figBCBepf} shows the 3$\sigma$ regions in the $\alpha$ vs.~$\gamma_2$ plane (marginalized over $D_{0xx}$) from the projected AMS-02 measurements of the \emph{(i)} proton flux (dashed lines), \emph{(ii)} B/C and $^{10}$Be/$^{9}$Be (dotted lines), and \emph{(iii)} B/C, $^{10}$Be/$^{9}$Be and proton flux (solid lines). Such as anticipated earlier, the ratios B/C and $^{10}$Be/$^{9}$Be are efficient probes of the diffusion index but insensitive to $\gamma_2$, while the proton flux constrains essentially the quantity $\alpha+\gamma_2$. Hence, within minimal propagation models, AMS-02 has the potential to pinpoint both the diffusion index $\alpha$ and the high energy source spectral index $\gamma_2$ with good accuracy. Note that this result stems exactly from the combination of the  B/C and $^{10}$Be/$^{9}$Be ratios and the proton flux. While a reasonably precise high energy proton flux has been measured by past and present instruments (e.g.~PAMELA \cite{pamelap}) $-$ allowing corrispondingly precise estimates of $\alpha+\gamma_2$ $-$, only AMS-02 (or future detectors) will be able to break down the degeneracy between $\alpha$ and $\gamma_2$ with high energy quality cosmic ray ratios.

\section{Conclusions}

\par In this paper, we have considered the ability of the upcoming AMS-02 experiment to measure stable secondary-to-primary and unstable ratios (such as boron-to-carbon and beryllium-10-to-beryllium-9), and studied to what extent this information could be used to constrain the model describing cosmic ray propagation in the Milky Way. Within the context of relatively simple propagation models, we find that the parameters can be very tightly constrained by the projected AMS-02 data; considerably more so than is possible with currently existing data~\cite{SimetHooper,DiBernardo:2009ku}.

The ability of AMS-02 to constrain the cosmic ray propagation model can be considerably reduced, however, if more complex models with larger numbers of free parameters are considered. Only a rough determination can be made of the Alfv\'en velocity (which dictates diffusive reacceleration) in most cases, for example. On the other hand, if even a relatively small degree of convection (at the level of $\sim$10 km/s/kpc) is present, this is expected to be discernable from the AMS-02 data. Other aspects of cosmic ray propagation (including, for example, the detailed energy dependence of the diffusion coefficient, or variations in the source distribution or injected chemical composition) are unlikely to be significantly constrained by upcoming data on stable secondary-to-primary and unstable ratios. In some cases, we have found that the parameters of the underlying diffusion model could be misinferred due to inaccurate assumptions implicit in the model. Local variations in recent supernova activity could also lead to somewhat skewed determinations of propagation parameters. The source spectral index, in constrast, is likely to be well constrained within minimal models if proton flux measurements are used in addition to the ratios B/C and $^{10}$Be/$^{9}$Be. With the latter case we have therefore exemplified how in some situations cosmic ray observables other than ratios are powerful tools in discriminating non-minimal assumptions.

In summary, the introduction of data from AMS-02 will make it possible to significantly expand our understanding of how cosmic rays propagate through the interstellar medium of the Milky Way. As we have demonstrated, the characteristics of simple propagation models will be tightly constrained by this data set. Moving beyond such simple scenarios, some of the underlying model assumptions will also become testable with such data, allowing us to better refine our cosmic ray predictions, and more generally expand our understanding of the Galactic cosmic ray spectrum.

{\it Acknowledgements:} DH is supported by the US Department of Energy, including grant DE-FG02-95ER40896, and by NASA grant NAG5-10842. MP acknowledges a grant from Funda\c{c}\~{a}o para a Ci\^encia e Tecnologia (Minist\'erio da Ci\^encia, Tecnologia e Ensino Superior) and thanks Gianfranco Bertone. MP also thanks the kind hospitality of the Center for Particle Astrophysics at Fermi National Accelerator Laboratory, where this project was started. All authors wish to thank the anonymous referee for the useful suggestions that helped improving the work.

\end{document}